\documentclass[letterpaper, 12pt]{article} 
\usepackage[margin=1in]{geometry}
\newcounter{somecounter}
\usepackage{booktabs, array}
\usepackage{setspace}
\usepackage{float}
\usepackage{epstopdf}
\usepackage{standalone}
\usepackage{color}
\usepackage{xcolor}
\usepackage{bm}
\usepackage{amsmath,amsfonts,amssymb}
\usepackage{amsthm}

\usepackage{pdflscape}
\usepackage{authblk}
\usepackage{graphicx}
\usepackage[colorlinks=true,citecolor=black,linkcolor=black,hidelinks]{hyperref}
\usepackage[normalem]{ulem}
\usepackage{multirow}
\usepackage{multicol}
\usepackage[flushleft]{threeparttable}
\usepackage{enumitem}
\usepackage{url}
\usepackage[utf8]{inputenc}
\usepackage{amsmath}
\usepackage{amsfonts}
\usepackage{amsthm}

\usepackage{lipsum}

\usepackage{natbib}
\newtheorem{theorem}{Theorem}    
 
\newtheorem{assumption}{Assumption}
\newtheorem{definition}{Definition}

\usepackage{mathtools}

\newcommand{\interior}[1]{%
 {\kern0pt#1}^{\mathrm{o}}%
}
\usepackage[english]{babel}
\usepackage[autostyle, english = american]{csquotes}
\MakeOuterQuote{"}

\usepackage{algorithm,algpseudocode}
\usepackage{caption}
\usepackage{subcaption}
\usepackage[makeroom]{cancel}

\makeatletter
\newcommand*\bigcdot{\mathpalette\bigcdot@{.5}}
\newcommand*\bigcdot@[2]{\mathbin{\vcenter{\hbox{\scalebox{#2}{$\m@th#1\bullet$}}}}}
\makeatother

\title{\textbf{Robust Nonparametric Testing Approaches for Spatial Regression}}

\author[1]{Kanghyun Wi}
\author[1]{Hyoeun Kim}
\author[3]{Tom\'a\v s Mrkvi\v cka}
\author[4]{Jorge Mateu}
\author[1,2]{Jaewoo Park}

\affil[1]{{Department of Statistics and Data Science}, {Yonsei University}}
\affil[2]{{Department of Applied Statistics}, {Yonsei University}}
\affil[3]{{Department of Data Science and Computing Systems}, {University of South Bohemia}}
\affil[4]{{Department of Mathematics}, {University Jaume I}}

\begin{document}

\def\spacingset#1{\renewcommand{\baselinestretch}%
{#1}\small\normalsize} \spacingset{1}

\maketitle

\begin{abstract}
Reliable inference for spatial regression remains challenging because it requires the correct specification of the spatial dependence structure, the mean trend, and the error distribution. Existing parametric testing methods rely on restrictive assumptions that are difficult to verify in practice and can lead to inaccurate conclusions under misspecification. To address this, we develop a robust nonparametric Monte Carlo testing framework for spatial regression based on random shifts. We construct test statistics that measure the dependence between residuals, obtained after removing the effects of nuisance covariates, and the covariate of interest. This allows us to assess the significance of the covariate in the sense of partial correlation. The proposed framework enables robust inference across various models without requiring parametric assumptions or even a closed-form distribution of the test statistics. Furthermore, we establish the asymptotic exactness of the random shift test in the increasing-domain setting when the sample covariance is used as the test statistic. Through extensive numerical experiments, we demonstrate that our method maintains the nominal significance level while achieving competitive power, whereas parametric methods can exhibit inflated type I error rates, even when they are correctly specified.
\end{abstract}

\noindent%
\vfill

\newpage
\spacingset{1.8} 

\section{Introduction}
\label{sec: Introduction}
Spatial regression models \citep[cf.][]{diggle1998model,cressie2015statistics} have been widely used to study the relationships between covariates and a response variable associated with spatial locations. Statistical inference in spatial regression analysis is an important problem with applications across disciplines such as climate science, ecology, and epidemiology. Although parametric approaches are commonly used, they are not robust to model misspecification. In this manuscript, we propose a robust Monte Carlo testing framework for spatial regression that allows both model estimation and hypothesis testing to be conducted in a fully nonparametric manner.

Parametric methods rely on several model assumptions. Typically, the mean function is assumed to be linear, and the spatial correlation is modeled using a stationary Gaussian process (GP) \citep{Banerjee2014, cressie2015statistics}, which imposes assumptions on the distribution and autocorrelation. By maximizing the likelihood function, one can estimate the model parameters and derive test statistics \citep{cressie2015statistics}. However, such model assumptions are difficult to verify in practice. If any of these assumptions are misspecified, parametric testing approaches can exhibit inflated type I error rates and lead to incorrect conclusions. In our studies, parametric methods exhibited empirical type I error rates between 0.1 and 0.2 under a nominal significance level of 0.05, and remained liberal even when they were correctly specified in finite-sample settings. In contrast, our method consistently maintained the nominal significance level. In addition, when the theoretical distribution of the test statistic is difficult to derive, classical parametric approaches can be limited.

Permutation tests, which assume exchangeability between observations under the null, can be more flexible because they do not require the closed-form distribution of the test statistic \citep{Winkler2014}. However, for spatial data, this assumption is violated because observations are correlated. Recently, \cite{RimalovaEtAl2022} proposed an iterative procedure to obtain exchangeable residuals and a Freedman–Lane permutation test \citep{Freedman1983} for spatial data. However, it still requires explicit assumptions about the mean trend and the covariance structure. Furthermore, the method involves repeated computations and can be numerically unstable depending on the model specification. This motivates the development of theoretically justified testing procedures that do not require restrictive model assumptions.

We develop a robust nonparametric inference framework for spatial regression. To remove the potential influence of nuisance covariates, we first obtain residuals by regressing the response on the nuisance covariates. This regression can be performed either parametrically \citep{cressie2015statistics, RimalovaEtAl2022} or nonparametrically \citep{Robinson2011, hastie1986generalized}. We then construct test statistics that capture the association between the residual field and the covariate of interest, allowing us to assess the significance of the covariate after adjusting for nuisance effects. Through random shift methods with either torus correction \citep{LotwickSilverman1982} or variance correction \citep{MrkvickaEtal2021b, Dvorak2024}, we perform Monte Carlo tests. These approaches do not rely on parametric assumptions for model estimation or derivation of the distribution of the test statistics.

We establish that the sample covariance test statistics of the random shift method with variance correction are asymptotically exchangeable under the null hypothesis (in the increasing-domain sense), ensuring asymptotic exactness of the test. We show that our method maintains the nominal significance level in settings where parametric approaches can become liberal, while achieving power comparable to existing methods across a variety of data-generating mechanisms.

The outline of this manuscript is as follows. Section~\ref{sec: Background} presents the background. In Section~\ref{sec: Methods}, we propose random shift methods for testing the significance of a covariate of interest while accounting for nuisance covariates. Section~\ref{sec: Theoretical results} provides a theoretical justification of the method. In Section~\ref{sec: Simulation Study}, we show extensive simulation studies comparing our nonparametric testing procedure with existing methods. The application of our approach to real spatial data examples comes in Section~\ref{sec: Application}. We finally conclude with a summary and discussion in Section~\ref{sec: Discussion}. 

\section{Background}
\label{sec: Background}

\subsection{Parametric Approaches for Spatial Regression}
\label{sec: Parametric Approaches for Regression Model}

Let $\mathcal{W} \subset \mathbb R^2$ be the compact spatial window of interest. For each spatial location $\mathbf{s} \in \mathcal{W}$, we have a response variable $y(\mathbf{s})$ and the covariates $x_1(\mathbf{s}),\dots,x_d(\mathbf{s})$. The parametric linear regression model is defined as 
\begin{equation}
\label{eq: parametric regression}
y(\mathbf{s}) = \beta_0 + x_1(\mathbf{s})\beta_1 + \cdots + x_d(\mathbf{s})\beta_d + \epsilon(\mathbf{s}),	
\end{equation}
where $\epsilon(\mathbf{s})$ follows a zero-mean GP. The GP is often modeled with the Mat\'{e}rn class \citep{stein2012interpolation} kernel with parameter $\bm{\xi}$. The standard approach to carry out inference is a maximum likelihood approach. Using the \texttt{nlme} package \citep{PinheiroBates2025}, we can estimate the parameters, $\bm{\xi},\bm{\beta} = (\beta_0,\dots, \beta_d)^\top$ and conduct significance tests.

Generalized least squares (GLS) approaches have also been widely used for inference for \eqref{eq: parametric regression}. Residuals can first be obtained using ordinary least squares estimates, and then, under the assumed covariance structure, $\bm\xi$ is estimated from these residuals by variogram fitting. The resulting covariance estimate is used to obtain the GLS estimate of $\bm{\beta}$. To assess the significance of GLS estimates, a permutation test has been studied \citep{HELWIG2019, DiCiccio2017}. The permutation test is a nonparametric procedure that assumes exchangeability. However, in the spatial regression, the exchangeability cannot be guaranteed due to the spatial dependence among the residuals. To approximate exchangeability, \cite{RimalovaEtAl2022} propose to decorrelate spatial residuals (i.e., filtering step) based on the estimated GLS covariance. Such a procedure is sensitive to model assumptions (e.g., mean trend and spatial correlation structure). When these assumptions are violated, the method tends to be overly liberal. Therefore, despite relying on permutations, this method still requires careful verification of multiple assumptions. In addition, the algorithm requires repeated covariance estimation via variogram fitting, which can be numerically unstable depending on the model specification.

\subsection{Nonparametric Approaches for Spatial Regression}
\label{sec: Nonparametric Approaches for Regression Model}

One can use a nonparametric regression to capture the nonlinear relationship as
\begin{equation}
	\label{eq: g_regression}
	y(\mathbf{s}) = f(x_1(\mathbf{s}),\dots,x_d(\mathbf{s})) + \epsilon(\mathbf{s}),
\end{equation}
where $f:\mathbb R^d \mapsto \mathbb R$ is the regression function and $\epsilon(\mathbf{s})$ is an error term. \cite{Robinson2011} proposed estimating the function $f(\cdot)$ using a Nadaraya–Watson (NW) kernel estimator, which assigns a higher weight to observations with more similar covariates. \cite{Robinson2011} studied the consistency and asymptotic normality of kernel estimates under sufficient conditions. We implement this approach by using the \texttt{npreg} function in the \texttt{np} package \citep{HayfieldRacine2008}. However, the convergence rate for estimating $f(\cdot)$ slows down considerably as $d$ increases \citep{hastie1986generalized}. To address this, generalized additive models (GAMs) have become a popular alternative for nonparametric spatial regression \citep[cf.][]{wikle2019spatio, yu2020estimation}. The model can be written as 
\begin{equation}
	\label{eq: gam}
	y(\mathbf{s}) = \sum_{j=1}^{d}f_j(x_j(\mathbf{s})) + \psi(\mathbf{s}) + \epsilon(\mathbf{s}),
\end{equation}
where $f_1(\cdot),\dots,f_d(\cdot)$ are univariate smooth functions for covariates, $\psi(\cdot)$ is a bivariate smooth function to account for spatial correlation, and $\epsilon(\mathbf{s})$ is an independent error term. In this manuscript, we investigate two types of GAMs: one in which the relationship between the covariates and the response (i.e., each $f_j(\cdot)$ function) is modeled parametrically, and another where this relationship is modeled nonparametrically. In both cases, $\psi(\cdot)$ is modeled nonparametrically, allowing flexibility without making explicit assumptions about the spatial covariance structure. We implement GAMs using the \texttt{gam} function in the \texttt{mgcv} package \citep{Wood2011}. We provide the detailed form of the nonparametric regression models in the supplementary material. 

\subsection{Random Shift Monte Carlo Tests}
\label{sec: Random shift Methods}

When the distribution of a test statistic is unknown, but replicated datasets can be generated under the null hypothesis, the Monte Carlo method is practical for hypothesis testing \citep{davison1997bootstrap}. The theoretical justification for this method relies on the permutation invariance of replicated test statistics under the null hypothesis, known as exchangeability. Exchangeability ensures the test achieves the nominal significance level under the null hypothesis.

The random shift method \citep{DaleFortin2002} is a nonparametric procedure for testing independence between two spatial processes by shifting one process relative to the other. Let $\Phi$ and $\Psi$ be two spatial processes over a compact observational window $\mathcal W$. Consider the case where $\Phi$ is fixed, and $\Psi$ is randomly shifted. The observed test statistic is computed from the data as $T_0 = T(\Phi, \Psi; \mathcal W)$. Denote $\Psi + \mathbf{v}_k$ a Euclidean shift of $\Psi$ by a random vector $\mathbf{v}_k$. Then, by applying random shift vectors $\mathbf{v}_1,\dots,\mathbf{v}_K$ to $\Psi$,  we obtain replicated statistics $T_k = T(\Phi, \Psi + \mathbf{v}_k; \mathcal W)$ for $k=1,\dots,K$. The Monte Carlo test is performed by assessing how extreme $T_0$ is relative to the entire set of replicated values $T_1,\dots, T_K$. Note that portions of the shifted process may fall outside the compact window $\mathcal W$, so we need correction methods to mitigate the distortion introduced by the shifts.

\subsubsection{Torus Correction}
\label{sec: Torus correction}

For a rectangular window $\mathcal{W}$, we can construct a toroidal geometry by gluing its opposite edges together \citep{LotwickSilverman1982}. Then, all observations can be used to generate replicates of the test statistics. We denote by $[\Psi + \mathbf{v}_k]$ the version of $\Psi$ shifted according to the toroidal geometry, in contrast to $\Psi + \mathbf{v}_k$. Then the replicates under the torus correction are computed as $T_k = T(\Phi, [\Psi + \mathbf{v}_k]; \mathcal{W})$ for $k=1,\dots, K$, where the random shift vectors $\mathbf{v}_k$ are sampled uniformly on $\mathcal{W}$.

Since the torus correction uses all observations, all test statistics $T_0,\ldots, T_K$ are directly comparable. However, toroidal shifts can introduce discontinuities in the autocorrelation structure of $\Psi$. Specifically, these “cracks” originate in areas that were originally distant in the data but become adjacent under the toroidal geometry created by gluing opposite edges. As a result, exchangeability is compromised, which can introduce biases in nominal significance levels, often leading to liberal tests.

\subsubsection{Variance Correction}
\label{sec: Variance correction}

\cite{MrkvickaEtal2021b} developed a variance-correction method that shifts the Euclidean window itself and discards any observations that fall outside it. This approach accommodates irregular observation windows and avoids the liberal bias introduced by toroidal shifts. However, since the amount of discarded observations varies across different random shift vectors $\mathbf{v}_k$, the variance of $T_k$ also fluctuates substantially, preventing the direct application of a Monte Carlo test. To address this issue, a priori standardization of $T_k$ is required.

Let $\mathcal W_k = \mathcal W \cap (\mathcal W + \mathbf{v}_k)$ be the intersection of the  window $\mathcal W$ and the shifted window $\mathcal W + \mathbf{v}_k$. After shifting both $\Psi$ and $\mathcal{W}$, we denote by $\Phi|_{\mathcal{W}_k}$ and $(\Psi + \mathbf{v}_k)|_{\mathcal{W}_k}$ the respective subsets that exclude any observations outside of $\mathcal{W}_k$. Then we can compute $T_k= T(\Phi|_{\mathcal W_k}, (\Psi + \mathbf{v}_k)|_{\mathcal W_k}; \mathcal W_k)$ for $k=1,\dots,K$. To make the replicates comparable, we standardize them to have zero mean and unit variance, $(T_k - \overline T) / \sqrt{\operatorname{var}(T_k)}$, and use the standardized statistics in the test. When an asymptotic formula for $\operatorname{var}(T_k)$ is available, we use it directly; otherwise, we can estimate $\operatorname{var}(T_k)$ via kernel regression \citep{MrkvickaEtal2021b}. 

\section{Methods}
\label{sec: Methods}

\subsection{Random Shift for Spatial Regression}   
\label{sec: Random shift for spatial regression}

We propose a fully nonparametric method for testing the significance of a covariate of interest while accounting for potential effects of nuisance covariates. Let $x_1(\mathbf{s}),\dots,x_d(\mathbf{s})$ denote the nuisance covariates and let $x_{d+1}(\mathbf{s})$ denote the covariate of interest. To account for potential dependence between the covariate of interest and the nuisance covariates, we first regress each $x_j(\mathbf{s})$ on $x_{d+1}(\mathbf{s})$, for $j=1,\ldots,d$. We then reconstruct the nuisance covariates as
\begin{equation}
    \label{eq: regress nuisance}
    \widetilde{x}_j(\mathbf{s}) =  \theta\cdot \widehat{g}(x_{d+1}(\mathbf{s})) + \delta(\mathbf{s}), \quad j=1,\dots, d,
\end{equation}
where $\widehat{g}(\cdot)$ is the fitted regression function capturing the dependence on the covariate of interest, $\delta(\mathbf{s})$ is the corresponding residual term, and $\theta\in [0,1]$ is a hyperparameter that controls the extent to which the dependence on $x_{d+1}(\mathbf{s})$ is retained in the reconstructed nuisance covariates. The choice of $\theta$ is discussed in detail below.

Subsequently, we regress the response $y(\mathbf{s})$ on the reconstructed covariates $\widetilde{x}_1(\mathbf{s}), \dots, \widetilde{x}_d(\mathbf{s})$ to remove the influence of the nuisance covariates on the response. The residuals are computed as
\begin{equation}
	\label{eq: residuals}
	e(\mathbf{s}) = y(\mathbf{s}) - \widehat{f}( \widetilde{x}_1(\mathbf{s}), \dots, \widetilde{x}_d(\mathbf{s}) ),
\end{equation}
where $\widehat{f}(\cdot)$ denotes the fitted mean trend as a function of the covariates. Note that both $\widehat{g}(\cdot)$ in \eqref{eq: regress nuisance} and $\widehat{f}(\cdot)$ can be specified either parametrically \eqref{eq: parametric regression} or nonparametrically \eqref{eq: g_regression}. Our method tests the null hypothesis that the residual process is independent of the covariate of interest.

Parameter $\theta$ in \eqref{eq: regress nuisance} induces a trade-off between type I and type II errors. As $\theta \to 0$, the component of the nuisance covariates explained by $x_{d+1}(\mathbf{s})$ is removed, retaining only the residual variation. Consequently, the residual term in \eqref{eq: residuals} retains more of the signal associated with $x_{d+1}(\mathbf{s})$, which may increase the type I error rate while reducing type II error. In contrast, as $\theta \to 1$, the original covariates are recovered, $\widetilde{x}_j(\mathbf{s}) = x_j(\mathbf{s})$, leading to improved control of type I error at the nominal level, potentially at the expense of increased type II error. In Section~\ref{sec: Simulation Study}, we observe that $\theta = 1$ controls the type I error at the nominal level while maintaining power comparable to the existing methods in most scenarios. We therefore suggest $\theta = 1$ as the default setting. However, users can select $\theta$ according to their specific objectives. Detailed simulation results are presented in Section~\ref{sec: Simulation Study}.

Suppose that we have observed locations $\mathbf{s}_1, \dots, \mathbf{s}_n$ over $\mathcal{W}$. Let $\mathbf{e}=(e(\mathbf{s}_1),\dots,e(\mathbf{s}_n))^\top$ be a realisation from the residual field $\Phi$, which is fixed, and let $\mathbf{x}_{d+1}=(x_{d+1}(\mathbf{s}_1),\dots,x_{d+1}(\mathbf{s}_n))^\top$ be a realisation from the random field $\Psi$ that will be randomly shifted. We consider both the torus and variance corrections described in the previous sections. For both methods, we assess whether the observed test statistic $T_0=T(\mathbf{e}, \mathbf{x}_{d+1};\mathcal W)$ is extreme relative to the replicate test statistics. In Section~\ref{sec: Test statistics}, we propose the test statistics that measure the correlation between the covariate of interest and the residuals obtained after removing the effects of the nuisance covariates. The entire procedure is described in Algorithms~\ref{alg: torus correction} and~\ref{alg: variance correction}.

\begin{algorithm}[htbp]
\caption{Random shift test for regression with torus correction}
\label{alg: torus correction}
\begin{algorithmic}
\State Following \eqref{eq: regress nuisance}, obtain the reconstructed $\widetilde{x}_1(\mathbf{s}), \dots, \widetilde{x}_d(\mathbf{s})$. 
\State Following \eqref{eq: residuals}, obtain the residuals from the fitted model $\widehat{f}(\cdot)$.
\State Compute the observed test statistic $T_0 = T(\mathbf{e}, \mathbf{x}_{d+1};\mathcal W)$.
\For{$k=1,\dots, K$}
\State Generate a vector $\mathbf{v}_k$ and shift $\mathbf{x}_{d+1}$ according to the toroidal geometry.
\State Compute the replicated statistics $T_k=T(\mathbf{e}, [\mathbf{x}_{d+1} + \mathbf{v}_k]; \mathcal W)$.
\EndFor
\State Determine whether the observed statistic is extreme relative to the shifted statistics at significance level $\alpha$.
\end{algorithmic}
\end{algorithm}

\begin{algorithm}[htbp]
\caption{Random shift test for regression with variance correction}
\label{alg: variance correction}
\begin{algorithmic}
\State Following \eqref{eq: regress nuisance}, obtain the reconstructed $\widetilde{x}_1(\mathbf{s}), \dots, \widetilde{x}_d(\mathbf{s})$. 
\State Following \eqref{eq: residuals}, obtain the residuals from the fitted model $\widehat{f}(\cdot)$.
\State Compute the observed test statistic $T_0 = T(\mathbf{e}, \mathbf{x}_{d+1};\mathcal W)$.
\For{$k=1,\dots, K$}
\State Generate a vector $\mathbf{v}_k$ and shift $\mathbf{x}_{d+1}$ according to the Euclidean geometry. 
\State Compute the replicated statistics $T_k= T(\mathbf{e}|_{\mathcal W_k}, (\mathbf{x}_{d+1} + \mathbf{v}_k)|_{\mathcal W_k}; \mathcal W_k)$.
\EndFor
\State Compute the standardized test statistics $(T_k - \overline T) /\sqrt{\operatorname{var}(T_k)}$ for $k=0,\dots, K$.
\State Determine whether the observed statistic is extreme relative to the shifted statistics at significance level $\alpha$.
\end{algorithmic}
\end{algorithm}

Our method is fully nonparametric in both the estimation and testing steps. It does not require specifying the form of $f(\cdot)$ and $g(\cdot)$ or making any assumptions about the residual structure. Users can simply fit $f(\cdot)$ and $g(\cdot)$ using their preferred models and then perform the random shift tests. This offers a major advantage over existing methods. For example, classical and permutation tests assume a parametric form of the regression function with stationary spatial correlation, which can lead to liberal behavior under model misspecification. In our study, we demonstrate that our nonparametric testing procedures achieve the nominal significance level more accurately than parametric procedures under model misspecification while maintaining comparable performance when the model is correctly specified.

\subsection{Test Statistics}
\label{sec: Test statistics}
By examining the dependence between the residuals, in which the effects of the nuisance covariates have been removed, and the covariate of interest, we can test the significance of $x_{d+1}(\mathbf{s})$. Here, we propose test statistics that can quantify the dependence between $x_{d+1}(\mathbf{s})$ and $e(\mathbf{s})$ obtained from \eqref{eq: residuals}. One natural choice for measuring their dependence is the sample covariance, defined as
\begin{equation}
    \nonumber
    \widehat C = \frac{1}{n-1}\sum_{i=1}^n \big(x_{d+1}(\mathbf{s}_i) - \overline{x}_{d+1}\big)\big(e(\mathbf{s}_i) - \overline{e}\big),
\end{equation}
where $\overline{x}_{d+1}$ and $\overline{e}$ denote the sample means of $x_{d+1}(\mathbf{s})$ and $e(\mathbf{s})$, respectively.

Since the above statistic mainly detects linear dependence, we additionally consider the sample distance covariance \citep{GaborRizzoBakirov2007}, which can capture general nonlinear dependence. Let $\mathrm{d}_{ij}(\mathbf e) = \| e(\mathbf s_i) - e(\mathbf s_j) \|_2$ denote the pairwise Euclidean distance between the residuals evaluated at locations $\mathbf s_i$ and $\mathbf s_j$. We define the centered distance metric for the residuals as
\begin{equation}
\begin{aligned}
\nonumber
\mathrm{D}_{ij}(\mathbf{e})
&= \mathrm{d}_{ij}(\mathbf{e})
- \frac{1}{n}\sum_{k=1}^n \mathrm{d}_{ik}(\mathbf{e})
- \frac{1}{n}\sum_{\ell=1}^n \mathrm{d}_{\ell j}(\mathbf{e})
+ \frac{1}{n^2}\sum_{k,\ell=1}^n \mathrm{d}_{k\ell}(\mathbf{e}) \\
&= \mathrm{d}_{ij}(\mathbf{e})
- \overline{\mathrm{d}}_{i.}(\mathbf{e})
- \overline{\mathrm{d}}_{.j}(\mathbf{e})
+ \overline{\mathrm{d}}_{..}(\mathbf{e}).
\end{aligned}
\end{equation}
$\mathrm{D}_{ij}(\mathbf{x}_{d+1})$ is defined analogously. The sample distance covariance is then given by
\begin{equation}
\nonumber
\mathrm{dCov}^2
= \frac{1}{n^2}\sum_{i,j=1}^n
\mathrm{D}_{ij}(\mathbf{e})\mathrm{D}_{ij}(\mathbf{x}_{d+1}).
\end{equation}

We use those statistics as $T_k$ for the random shift methods. Note that the variance correction requires standardizing the test statistics, which in turn requires an estimate of $\operatorname{var}(T_k)$. When we have $n$ observations, \cite{MrkvickaEtal2021b} showed that the asymptotic variance order of $\widehat C$ follows $1/n$ for two general spatial random fields. Thus, it also holds for the proposed residual field and the covariate of interest. In addition, \citet{GaborRizzoBakirov2007} shows that, under the null hypothesis of independence, the normalized statistic 
\[
\frac{n\,\mathrm{dCov}^2}{
\overline{\mathrm{d}}_{..}(\mathbf{e})\,
\overline{\mathrm{d}}_{..}(\mathbf{x}_{d+1})
}
\]
converges in distribution to a quadratic form of standard Gaussian variables. Motivated by this asymptotic behavior, we use $1/n_k$ as an estimate of $\operatorname{var}(T_k)$ for the statistics $\widehat{C}$, where $n_k$ denotes the number of observations in the intersection window $\mathcal{W}_k$. For $\mathrm{dCov}^2$, we instead rescale the statistic by $\{\overline{\mathrm{d}}_{..}(\mathbf{e})\overline{\mathrm{d}}_{..}(\mathbf{x}_{d+1})\}/n_k$.

\subsection{Nonparametric Variable Selection}
\label{sec: Nonparametric Variable Selection}

Although the primary goal of our framework is hypothesis testing for covariate significance rather than variable selection in the broad sense, the proposed procedure can also be naturally extended to backward stepwise variable selection (Algorithm~\ref{alg: stepwise inference}). Let $\mathbf{x}_{j}=(x_{j}(\mathbf{s}_1),\dots,x_{j}(\mathbf{s}_n))^\top$ denote the $j$-th covariate and $\mathbf{x}_{-j}=(\mathbf{x}_{1},\dots,\mathbf{x}_{j-1},\mathbf{x}_{j+1},\dots, \mathbf{x}_{d+1})^\top$ denote the set of all remaining covariates except $\mathbf{x}_j$. We start by performing a random shift test on each $\mathbf{x}_j$, treating it as the covariate of interest while considering the remaining covariates $\mathbf{x}_{-j}$ as nuisance variables. We then compute the $p$-values $\widehat{p}_j$ for each $j$. Next, we iteratively remove the covariate with the highest $p$-value and repeat this procedure until the $p$-values for the remaining covariates are smaller than $\alpha$. 

\begin{algorithm}[htbp]
\caption{Nonparametric variable selection for spatial regression}
\label{alg: stepwise inference}
\begin{algorithmic}
\Repeat
\For{each $j$}
\State Let $\mathbf{x}_{j}$ be a covariate of interest and $\mathbf{x}_{-j}$ be the nuisance covariates. 
\State Following \eqref{eq: regress nuisance}, obtain $\widetilde{\mathbf{x}}_{-j}$.
\State Following \eqref{eq: residuals}, obtain the residuals from the fitted model $\widehat{f}(\widetilde{\mathbf{x}}_{-j})$.
\State Following Algorithm \ref{alg: torus correction} or \ref{alg: variance correction}, compute $\widehat p_j$.
\EndFor
\If{$\max_j \widehat{p}_j >\alpha$}
\State Remove the covariate with the highest $p$-value.
\EndIf
\Until the $p$-values for the remaining covariates are smaller than $\alpha$.
\end{algorithmic}
\end{algorithm}

\section{Theoretical Results}
\label{sec: Theoretical results}

To guarantee exactness, permutation tests rely on an exchangeability assumption \citep{Winkler2014}, which is difficult to justify in the presence of spatial dependence because the observations are no longer invariant under permutation \citep{Westfall1993}. In this section, we show that the sample covariance test statistic is asymptotically exchangeable under the null hypothesis, and that the random shift test with a variance correction is asymptotically exact.

\subsection{Asymptotic Exactness of the Test}
Let $\Phi(\mathbf{s})$ and $\Psi(\mathbf{s})$, $\mathbf{s} \in \mathbb{R}^2$, be stationary zero-mean random fields that are independent under the null hypothesis. They can represent the residual and covariate processes in our case. Stationarity ensures that their joint distributions are invariant under random shifts, which is essential for our results. We consider an increasing-domain asymptotic regime, described by \cite{Lahiri}, in which observations $\mathbf s_{1},\dots, \mathbf s_{n}$ are collected over a sequence of increasing windows $\{\mathcal{W}^n\}_{n\ge 1}$. Define the observational window $\mathcal W^0$ as a bounded Borel subset of $\mathbb{R}^2$ containing the origin. Let $\mathbf u_{1},\dots,\mathbf u_{n}$ be locations independently sampled from a distribution with a continuous and strictly positive density $p(\mathbf{u})$ on the closure of $\mathcal{W}^0$. The increasing windows $\lbrace \mathcal W^n \rbrace_{n\geq 1}$ and the locations $\mathbf s_{1},\dots,\mathbf s_{n}$ are obtained by scaling $\mathcal W^0$ and $\mathbf u_{1},\dots,\mathbf u_{n}$ by a multiplicative factor. Following \cite{Lahiri}, we index both the windows and the locations by $n$ for notational convenience, although the indices need not be the same. This is formalized in the following assumption.

\begin{assumption}\label{asm: windows and boundaries}
	Let $\{\lambda_n\}_{n\in\mathbb N}$ be a sequence of positive real numbers such that $n^{\epsilon}/\lambda_n\rightarrow 0$ as $n\rightarrow \infty$ for some $\epsilon > 0$. Then, for each $n$,
    \[
    \mathcal W^n = \lambda_n \mathcal W^0,\quad 
    \mathbf s_{i} = \lambda_n \mathbf u_{i},\quad 1 \le i \le n.
    \]
Furthermore, for any sequence $\{a_n\}$ with $a_n \to 0+$, let 
$\{a_n(\mathbf z + [0,1)^2) : \mathbf z \in \mathbb Z^2\}$ 
be the associated collection of cubes. Then the number of cubes 
that intersect both $\mathcal W^0$ and $(\mathcal W^0)^c$ is 
$\mathcal O(1/a_n)$ as $n \to \infty$. 
\end{assumption}

Assumption~\ref{asm: windows and boundaries} is a standard regularity condition controlling the complexity of the boundary. Because $\mathcal W^n = \lambda_n \mathcal W^0$, the shape of the observation window remains unchanged with $n$. \cite{ShermanCarlstein} showed that the assumption holds when the boundary of $\mathcal W^0$ is smooth, allowing for fairly general window shapes.

In variance correction, we define the intersection windows $\mathcal{W}^n_k = \mathcal{W}^n \cap (\mathcal{W}^n + \mathbf{v}_k)$,  for each shift vector $\mathbf{v}_k$. For each $\mathcal{W}^n_k$, the corresponding statistic is
$T_k^n = T(\Phi|_{\mathcal{W}^n_k}, (\Psi+\mathbf{v}_k)|_{\mathcal{W}^n_k}; \mathcal{W}^n_k)$ and its standardized version $(T_k^n-\overline{T^n})/\sqrt{\mathrm{Var}(T_k^n)}$. The collection $(T_0^n, \ldots, T_K^n)$ is defined to be asymptotically exchangeable as follows.

\begin{definition}\label{def: asymptotic exchangeability}
The sequence of test statistics $(T_0^n, \ldots, T_K^n) \in \mathcal{M}^{K+1}$, where $\mathcal{M}$ is a measurable space, is said to be asymptotically exchangeable as $n\rightarrow\infty$, if its limiting joint distribution is invariant under permutations, i.e., 
$$\lim_{n\rightarrow\infty}\big[\mathbb{P}\{(T_0^n, \ldots, T_K^n) \in A\}-\mathbb{P}\{(T_{\sigma (0)}^n, \ldots , T_{\sigma (K)}^n) \in A\}\big]=0$$
for any measurable set $A \subset \mathcal{M}^{K+1}$ and any permutation $\sigma$.
\end{definition}

Definition~\ref{def: asymptotic exchangeability} is key to permutation tests, as it provides a sufficient condition for their asymptotic exactness. Formally, the test is asymptotically exact as follows.

\begin{theorem}\label{thm: AE}
Let $(T_0^n, \ldots, T_K^n) \in \mathcal{M}^{K+1}$ be asymptotically exchangeable test statistics, and let $1(T_k^n < T_0^n)$ be the indicator variable that equals one if $T_k^n < T_0^n$. Then the sequence of Monte Carlo tests with $p^n=\frac{1}{K+1} (1+\sum_{k=1}^K 1 (T_k^n<T_0^n))$ is asymptotically exact, i.e., 
$$\lim_{n\rightarrow\infty}\mathbb{E}\{1 (p^n\leq \alpha )\} = \alpha,$$ provided that $\alpha (K+1)$ is an integer.
\end{theorem}

Proof of Theorem~\ref{thm: AE} comes in the supplementary material. By Theorem~\ref{thm: AE}, it suffices to establish the asymptotic exchangeability of $(T_0^n, \ldots, T_K^n)$ to show asymptotic exactness of the Monte Carlo test. One way to achieve asymptotic exchangeability is to show that these statistics are asymptotically independent and identically distributed. In what follows, we show this for the sample covariance test statistics.

\subsection{Asymptotic Exchangeability of the Test Statistics}

Let ${T}^n_k = {n_k}^{-1} \sum_{\mathbf{s}_{i}\in \mathcal{W}^n_k} \Phi(\mathbf{s}_{i})\Psi(\mathbf{s}_{i} -\mathbf{v}_k)$ 
be the sample covariance between $\Phi$ and $\Psi$ over $\mathcal{W}^n_k$, where $n_k$ is the number of points in $\mathcal W^n_k$. In practice, the estimator is centered and scaled by $1/(n_k-1)$. These adjustments affect only finite samples and do not change the asymptotic distribution; hence, we use the uncentered version with scaling $1/n_k$ for notational simplicity. To derive the asymptotic distribution of ${\mathbf{T}}^n = ({T}^n_0,\dots,{T}^n_K)$ under the null, we introduce the following assumptions.

\begin{assumption}\label{asm: alpha mixing}
The random fields $\Phi(\mathbf s)$ and $\Psi(\mathbf s)$ are strongly mixing with coefficient $\alpha(a;b)$ defined in \cite{Lahiri}, where $a>0$ denotes the separation distance between sets and $b>0$ their size. 
Suppose that $\alpha(a;b) \leq \alpha_1(a)h(b)$, where $\alpha_1(\cdot)$ is nonincreasing with $\lim_{a\rightarrow\infty}\alpha_1(a)=0$ and $h(\cdot)$ is nondecreasing.
\end{assumption}

\begin{assumption}\label{asm: mixing bound}
There exist $\{\lambda_{1n}\}$, $\{\lambda_{2n}\}$ with $\lambda_{1n}\geq \lambda_{2n}\geq \log\lambda_n$, such that
\begin{equation}
    \begin{aligned}
        \nonumber
        &(\log n)^2 (\lambda_n^{-1}\lambda_{1n} + \lambda^{-1}_{1n}\lambda_{2n})=o(1),  \\
        &(\log n)^4 \lambda_n^{-2}\lambda_{1n}^2  \sum_{q=1}^{\lambda_{1n}}q^3\alpha_1(q)=o(1),  \\
        &\lambda_{1n}^{-2}\lambda_n^2\alpha_1(\lambda_{2n})h(\lambda_n^2) =o(1),  \\
        &h(\lambda_{1n}^2)\{\lambda_{1n}^2\alpha_1(\lambda_{2n}) + \sum_{q=\lambda_{1n}}^\infty q\alpha_1(q)\}=o(1). 
    \end{aligned}
\end{equation}
\end{assumption}

\begin{assumption}\label{asm: covariance existence} For some $\delta > 0$, assume that
	\begin{equation}
		\nonumber
		\mathbb E|\Phi(\mathbf 0)|^{2+\delta}<\infty,\; \mathbb E|\Psi(\mathbf 0)|^{2+\delta}<\infty\text{ and }\int_1^\infty t\alpha_1(t)^{\frac{\delta}{2+\delta} }dt<\infty.
	\end{equation}
\end{assumption}

\begin{assumption}\label{asm: compact support}
Assume that the random fields $\Psi$ and $\Phi$ have compactly supported covariance; that is, there exists $R>0$ such that
$\mathbb{E}\left[\Psi(\mathbf{0})\Psi(\mathbf{h})\right] = 0$
for all $\|\mathbf{h}\|>R$, and the same holds for $\Phi$.
\end{assumption}

\begin{assumption}\label{asm: shift vectors}
Assume that there exist fixed non-zero vectors $\mathbf{v}_1, \dots, \mathbf{v}_K$ such that $\|\mathbf{v}_k\| > 2R$, and $\|\mathbf{v}_k - \mathbf{v}_l\| > 2R$ for $k \neq l$, where $k,l=1,\dotsm, K$.
\end{assumption}

Assumption~\ref{asm: alpha mixing} implies that the dependence between the random fields weakens as the lag increases, so local dependence becomes negligible.
Assumption~\ref{asm: mixing bound} provides sufficient conditions on the strong mixing coefficient, and Assumption~\ref{asm: covariance existence} ensures the existence of the limiting covariance. Assumptions~\ref{asm: windows and boundaries}--\ref{asm: covariance existence} follow \cite{Lahiri} and are standard in the literature on increasing domain asymptotics. Assumption~\ref{asm: compact support} covers compactly supported covariance functions commonly used in large spatial domains \citep{Wendland, Wu}. Assumption~\ref{asm: shift vectors} guarantees asymptotic independence of test statistics at different shifts. Here, the shift vectors are fixed and do not grow with $n$. It is not restrictive in practice, as suitable fixed shift vectors can be chosen when the spatial domain is sufficiently large. Under these assumptions, we establish the following result.

\begin{theorem}\label{thm: AN} 
Let $\mathbf D_n=\operatorname{diag}(n_0,\dots,n_K)$ be the diagonal matrix in which each diagonal element represents the number of points in $\mathcal{W}^n_k$. We denote $C_\Phi(\cdot)$ and $C_\Psi(\cdot)$ as the autocovariance functions of the random fields $\Phi$ and $\Psi$, respectively. Suppose that Assumptions~\ref{asm: windows and boundaries}, \ref{asm: alpha mixing}, \ref{asm: mixing bound}, \ref{asm: covariance existence}, \ref{asm: compact support} and~\ref{asm: shift vectors} hold. Under the null hypothesis, if 
$n/\lambda_n^2\rightarrow C_1\in (0,\infty)$ as $n\rightarrow\infty$, then
\begin{equation}
    \nonumber
    \mathbf D_n^{1/2}{\mathbf{T}}^n \overset{d}{\to}\mathcal{N}_{K+1}(\mathbf{0}, \sigma^2\mathbf{I})
    \quad \text{as } n \to \infty,
\end{equation}
where $\sigma^2=C_\Phi(\mathbf{0}) C_\Psi(\mathbf{0}) + C_2 \int_{\mathbb{R}^2} C_\Phi(\mathbf{h}) C_\Psi(\mathbf{h}) d\mathbf{h}$, and $C_2=C_1\int_{\mathcal{W}^0} p^2(\mathbf{u})d\mathbf{u}.$

\end{theorem}

Proof of Theorem~\ref{thm: AN} is in the supplementary material. Theorem~\ref{thm: AN} shows that, under the null hypothesis, the sample covariance test statistics are asymptotically independent and identically distributed, and hence asymptotically exchangeable in the sense of Definition~\ref{def: asymptotic exchangeability}. Theorem~\ref{thm: AN} can be applied directly to the regression setting with $\Phi(\mathbf{s}) = e(\mathbf{s})$ defined in \eqref{eq: residuals} and $\Psi(\mathbf{s}) = x_{d+1}(\mathbf{s})$, the covariate of interest. Therefore, if $e(\mathbf{s})$ and $x_{d+1}(\mathbf{s})$ satisfy the conditions of Theorem~\ref{thm: AN}, the resulting test statistics are asymptotically exchangeable under the null. Together with Theorem~\ref{thm: AE}, this implies that the random shift test with variance correction and sample covariance as the test statistic is asymptotically exact.

\section{Simulation Study}
\label{sec: Simulation Study}

We fit four models— a spatial linear model, an NW kernel estimator, and GAMs with linear and nonlinear mean structures—denoted by LM, NW, $\mbox{GAM}_{L}$, and $\mbox{GAM}_{NL}$, with details provided in the supplementary material. For LM, we conduct the significance tests either using the classical $p$-value test \citep{cressie2015statistics} or by using the permutation test \citep{RimalovaEtAl2022}. For the permutation test, parameters are estimated via iterative GLS following \cite{RimalovaEtAl2022}. For the nonparametric models (i.e.,  NW, $\mbox{GAM}_{L}$, and $\mbox{GAM}_{NL}$), we apply random shift methods with variance corrections. Given their quantitative similarity, results based on Kendall’s correlation \citep{Dvorak2024} and torus correction are reported in the supplementary material. 

\subsection{A Single Nuisance Covariate with Various Spatial Error Structures}
\label{sec: Scenarios for single nuisance covariate}
\begin{figure}[htbp]
\centerline{\includegraphics[width=10cm]{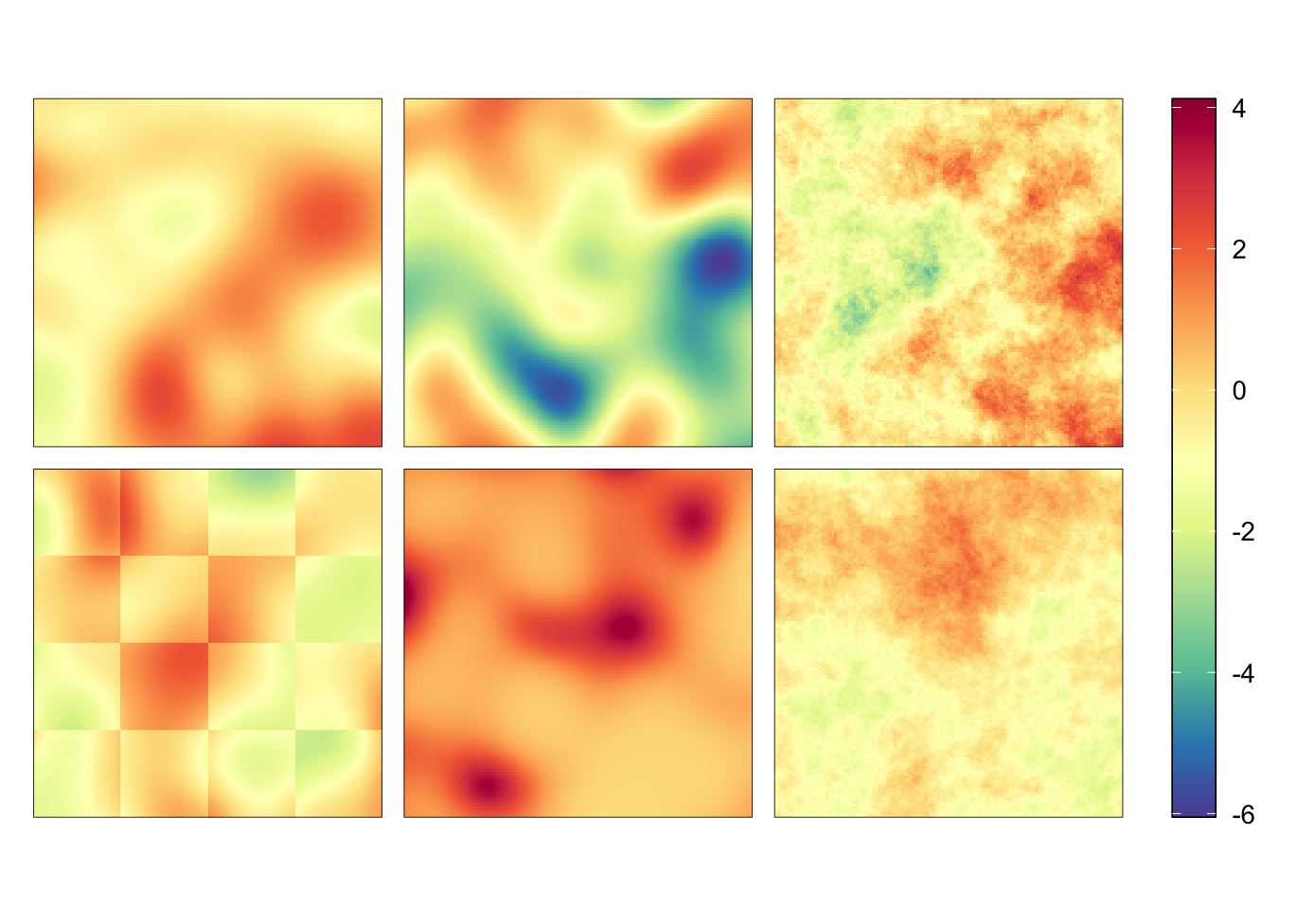}}
\caption{Simulated spatial random fields under different error structures. From left to right, top to bottom: SE1, SE4, E1, N, LN, and NS.}
\label{simulscenario}
\end{figure}

We study the liberality of the tests under a single-nuisance-covariate case. We generate the observed locations $\mathbf{s}_1, \ldots, \mathbf{s}_{100}$ uniformly over $[0 ,1]^2$. The nuisance covariate $x_1(\mathbf{s})$ and the covariate of interest $x_2(\mathbf{s})$ are independently simulated from a GP with mean zero and an exponential covariance function with variance $1$ and range $0.2$. To assess robustness under model misspecification, we considered a variety of spatial error structures for $\epsilon(\mathbf{s})$, which are described below and illustrated in Figure~\ref{simulscenario}.

\begin{enumerate}
    \item (SE1) GP with squared exponential covariance, variance $1$, range $0.2$, and nugget $1$.
    \item (SE4) Same as SE1 with variance $4$.
    \item (E1) GP with exponential covariance, variance $1$, range $0.2$, and nugget $1$.
    \item (N) GP with squared exponential covariance and a spatially alternating mean over $16$ subwindows, inducing checkerboard-like negative dependence.
    \item (LN) Log-GP whose logarithm follows SE1.
    \item (NS) Nonstationary GP generated via kernel convolution \citep{Mark2017}, with kernels identical to E1.
\end{enumerate}
Details of the random error structures are provided in the supplementary material. We test the null hypothesis that $y(\mathbf{s})$ and $x_2(\mathbf{s})$ are independent given $x_1(\mathbf{s})$ in all scenarios. For the linear mean trend case, $y(\mathbf{s})$ is generated under $y(\mathbf{s})=-0.5 + x_1(\mathbf{s})  + \epsilon(\mathbf{s})$. For the nonlinear mean trend case, $y(\mathbf{s})$ is generated under $y(\mathbf{s})=-0.5 + x^2_1(\mathbf{s})  + \epsilon(\mathbf{s})$. We study whether each test method can achieve the nominal 0.05 significance level. Specifically, for each scenario (6 error structures $\times$ 2 mean structures = 12 total), we simulate data under the null hypothesis. 

\begin{table}[htbp]
\caption{Violated assumptions by data-generating scenario (“M”: mean, “C”: correlation, “D”: distribution).}
\centering
\scriptsize               
\setlength{\tabcolsep}{4pt} 
\renewcommand{\arraystretch}{1.0} 
\begin{tabular}{c c *{6}{c}}
\toprule
  \multirow{2}{*}{\textbf{Test (model)}} & \multirow{2}{*}{\textbf{Mean}} &
  \multicolumn{6}{c}{\textbf{Error}} \\
  & & SE1 & SE4 & E1 & N & LN & NS \\
\hline
  Classical method (LM)                & \multirow{5}{*}{Linear}    & - & - & C & C & D & C \\
  Permutation test (LM)                &                                     & - & - & C & C & D & C \\
  Random shift ($\mathrm{GAM}_L$) &                                     & - & - & -   & -   & D & -   \\
  Random shift (NW)               &                                     & - & - & -   & -   & -    & -   \\
  Random shift ($\mathrm{GAM}_{NL}$)&                                  & - & - & - & - & D & -    \\
\midrule
  Classical method (LM)                & \multirow{5}{*}{Nonlinear} & M & M & M / C & M / C & M / D & M / C \\
  Permutation test (LM)                &                                     & M & M & M / C & M / C & M / D & M / C \\
  Random shift ($\mathrm{GAM}_L$) &                                     & M & M & M & M & M / D & M \\
  Random shift (NW)               &                                     & - & - & - & - & - & - \\
  Random shift ($\mathrm{GAM}_{NL}$)&                                   & - & - & - & - & D & - \\
\bottomrule
\end{tabular}
\label{viol_assump}
\end{table}

Here, we assume a squared-exponential covariance for LM. Consequently, model misspecification arises in the four scenarios that do not follow this covariance structure (i.e., all scenarios except SE1 and SE4). Furthermore, when the true mean structure is nonlinear, the LM is also violated. Therefore, model misspecification often arises in classical methods and permutation tests that rely on LM fitting. On the other hand, random shift tests, which are based on nonparametric models (NW, $\mbox{GAM}_{L}$, and $\mbox{GAM}_{NL}$), exhibit fewer model misspecifications. Table~\ref{viol_assump} summarizes the violated assumptions for different test methods under various scenarios.

\begin{figure}[htbp]
\centerline{\includegraphics[width=\textwidth]{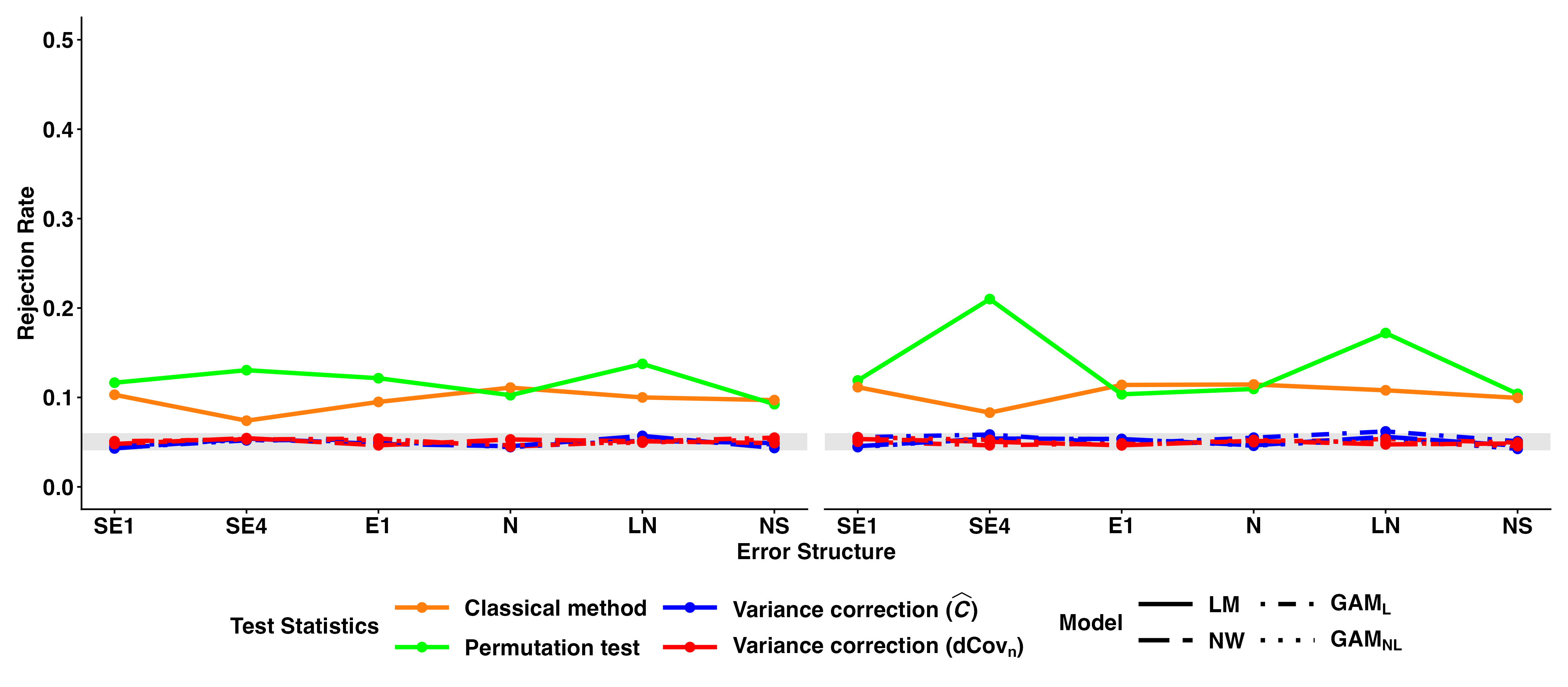}}
\caption{Empirical rejection rates under the linear model (left) and nonlinear model (right). The shaded horizontal band indicates the 95\% binomial confidence interval for the rejection rate at $\alpha=0.05$, given by [0.041, 0.060].}
\label{siglev}
\end{figure}

For the random shift methods, we set $\theta =1$, the default setting. We generated 499 Monte Carlo samples of the test statistics, which were then compared with the observed test statistics to compute the $p$-value. We repeat this process 2,000 times and count how often the null hypothesis is incorrectly rejected. We assess the liberality of the different methods by comparing the rejection rates with the 95\% confidence interval of the binomial distribution with 2,000 trials and a success probability of $0.05$ (i.e., the significance level). Figure~\ref{siglev} summarizes the results.

Figure~\ref{siglev} highlights that our nonparametric approaches are robust to the choice of model and test statistic. The variance correction keeps rejection rates within the nominal confidence interval across various spatial error structures and trend models. Both the classical method and the permutation test exhibit liberality across all scenarios. Although classical methods and permutation tests are expected to match the nominal significance level under the correct covariance structure, they exhibit liberality for SE1 and SE4 in both trend scenarios. In the supplementary material, we observe that the classical test becomes close to the nominal rate with increasing sample size, but remains liberal in finite samples. In contrast, the permutation test suffers from numerical instability during the iterative variogram fitting and remains liberal even for large samples. Under model misspecification (i.e., E1, N, LN, NS), both the classical method and the permutation test consistently exhibit substantial liberality. These observations indicate that our nonparametric approaches exhibit the most stable and conservative behavior across different scenarios. In the supplementary material, we present power analyses showing that our nonparametric tests typically have slightly lower but comparable power than the classical methods. Since classical methods are liberal, their higher power is expected. Under nonlinear mean trends, however, using distance covariance as the test statistic leads to higher power than the classical methods, as it captures nonmonotone dependence.

\subsection{Multiple Independent Covariates}
\label{sec: Multiple Independent Covariates with Varying theta}

In this section, we consider the multiple regression setting. The locations $\mathbf{s}_1,\ldots,\mathbf{s}_{100}$ are sampled uniformly over $[0 ,1]^2$. We simulate four covariates, $x_1(\mathbf{s}),\ldots,x_4(\mathbf{s})$, independently from a zero-mean GP with distinct covariances, as described below.

\begin{enumerate}
    \item $x_1(\mathbf{s})$: GP with an exponential covariance, unit variance, and range $0.2$.
    \item $x_2(\mathbf{s})$: GP with a stable covariance, unit variance, shape 0.5, and range 0.1. 
    \item $x_3(\mathbf{s})$: GP with a Mat\'ern covariance, unit variance, smoothness 2, and range 0.1.
    \item $x_4(\mathbf{s})$: GP with a generalized Cauchy covariance, unit variance, smoothness $\alpha=2$, tail parameter $\beta=5$, and range $0.1$.
\end{enumerate}
We generate the response $y(\mathbf{s})$ under two models: a linear trend, $y(\mathbf{s}) = -0.5+ x_1(\mathbf{s}) + x_2(\mathbf{s}) +x_4(\mathbf{s})+\epsilon(\mathbf{s})$ and a nonlinear trend, $y(\mathbf{s}) = -0.5+ \exp\{x_1(\mathbf{s})\} + x_2^2(\mathbf{s}) +x_4^3(\mathbf{s})+\epsilon(\mathbf{s})$. The error process $\epsilon(\mathbf{s})$ follows a GP with a squared exponential covariance, consistent with the SE1 setting in Section~\ref{sec: Scenarios for single nuisance covariate}. To examine the effect of $\theta$, we consider several values of $\theta$ in the backward variable selection procedure with variance correction, including $0$, $0.25$, $0.5$, $0.75$, and $1$. For each scenario (2 mean trends $\times$ 5 values of $\theta$, for a total of 10), we evaluate type I error control for $x_3(\mathbf{s})$, which is not included in the data-generating model, and empirical power for $x_1(\mathbf{s})$, $x_2(\mathbf{s})$, and $x_4(\mathbf{s})$. $\theta$ applies only to our nonparametric approaches and is not defined for the classical method or permutation test.

\begin{figure}[htbp]
\centerline{\includegraphics[width=\textwidth]{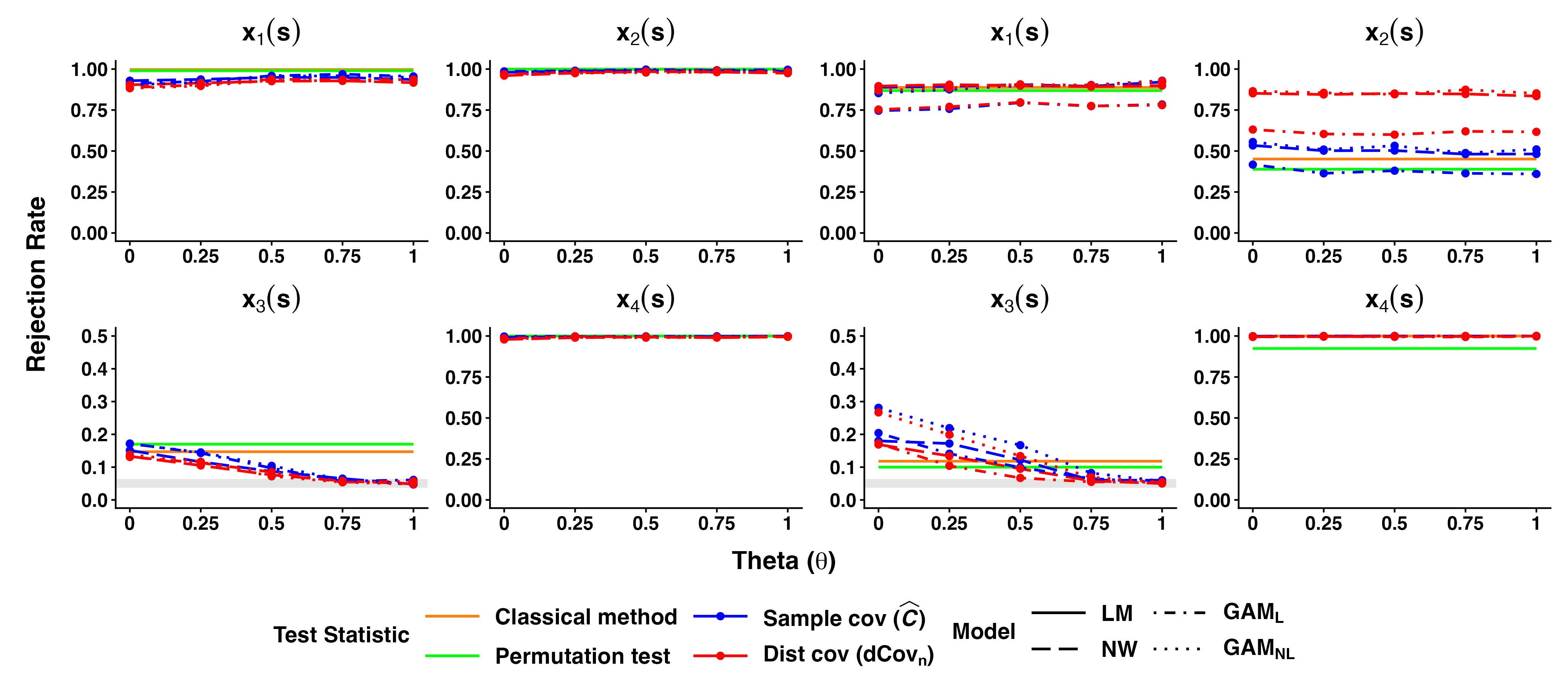}}
\caption{Empirical rejection rates under the linear model (left) and the nonlinear model (right). The shaded horizontal band indicates the 95\% binomial confidence interval for the rejection rate at $\alpha=0.05$, given by [0.037, 0.064].}
\label{fig: s1}
\end{figure}

For the random shift methods and the permutation test, we generate 499 Monte Carlo samples of the test statistic. We repeat this procedure 1,000 times, computing the empirical rejection rate of each covariate at the conclusion of the backward variable selection. For the classical $t$-test, backward selection is repeated 1,000 times using parametric $p$-values, without Monte Carlo simulation. For $x_3(\mathbf{s})$, we assess liberality by comparing the rejection rate to the 95\% binomial confidence interval with 1,000 trials and success probability $0.05$. The results are summarized in Figure~\ref{fig: s1}.

Overall, the classic and permutation tests exhibit inflated type I error rates, whereas our nonparametric approaches achieve nominal significance at $\theta=1$ under both linear and nonlinear trends. In the linear case, the random-shift-based methods achieve power comparable to the classical and permutation tests for $x_1(\mathbf{s})$, $x_2(\mathbf{s})$, and $x_4(\mathbf{s})$ across all values of $\theta$. In the nonlinear case, the sample covariance yields power similar to the classical and permutation tests, while the distance covariance achieves higher power, particularly for $x_2(\mathbf{s})$, highlighting its advantage in capturing nonmonotone relationships. Across both mean trends, decreasing $\theta$ from 1 to 0 leads to increased type I error with little change in type II error. These results imply that $\theta = 1$ achieves nominal type I error control with comparable power.

\subsection{Multiple Dependent Covariates}
\label{sec: Multiple Dependent Covariates with Varying theta}

We now consider dependent covariates under the same simulation design as in Section~\ref{sec: Multiple Independent Covariates with Varying theta} but introduce dependence between $x_1(\mathbf{s})$ and $x_4(\mathbf{s})$. Here, $x_4(\mathbf{s})$ is generated as a GP with mean proportional to $x_1(\mathbf{s})$ and the same covariance. For the random shift methods and the permutation test, we generate 499 Monte Carlo samples of the test statistic and repeat the entire procedure 1,000 times.

\begin{figure}[htbp]
\centerline{\includegraphics[width=\textwidth]{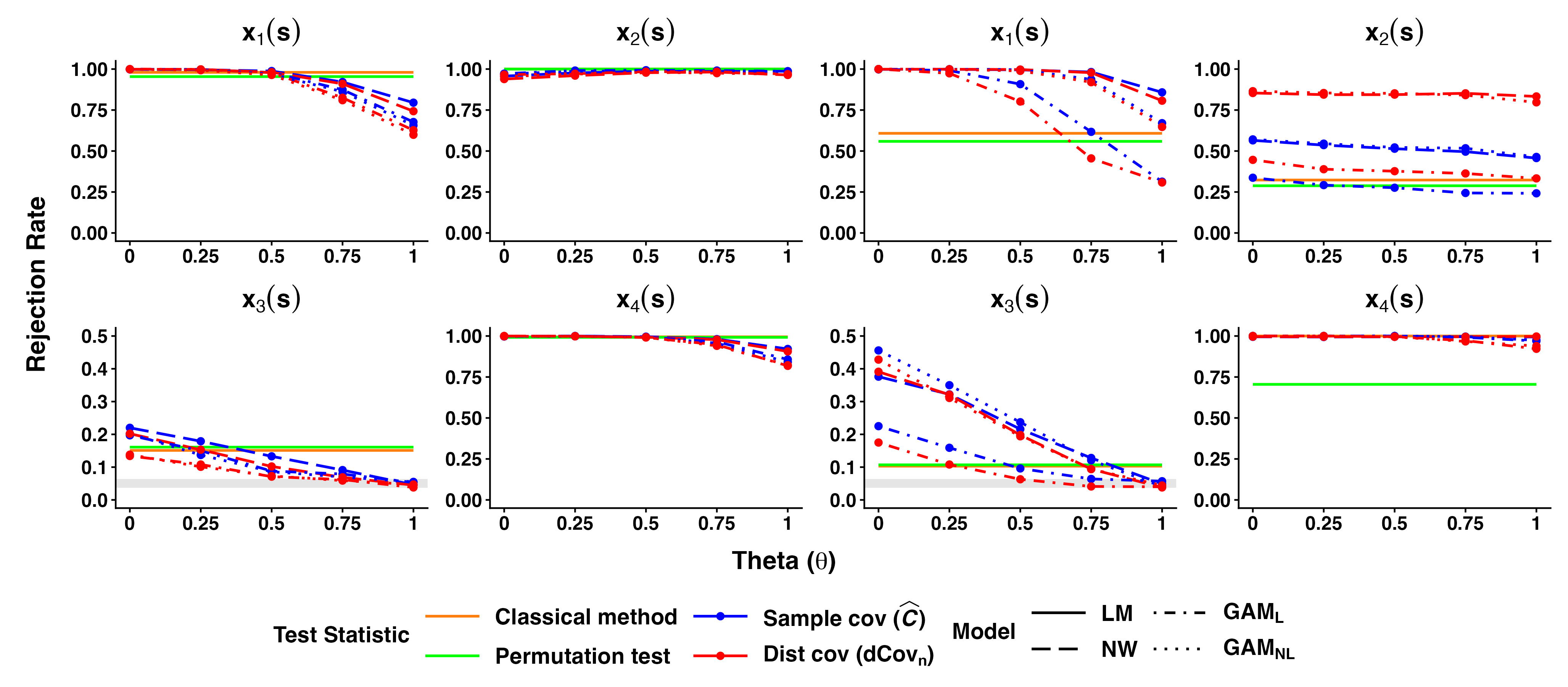}}
\caption{Empirical rejection rates under the linear model (left) and the nonlinear model (right). The shaded horizontal band indicates the 95\% binomial confidence interval for the rejection rate at $\alpha=0.05$, given by [0.037, 0.064].}
\label{fig: s2}
\end{figure}

Figure~\ref{fig: s2} shows that, with $\theta=1$, the random shift approaches achieve the nominal rate. In contrast, the classical and permutation tests exhibit liberality for $x_3(\mathbf{s})$, as in the previous examples. For the random shift approaches, there is a trade-off between type I and type II errors as $\theta$ varies. Specifically, as $\theta \to 1$, our methods achieve nominal type I error control at the expense of reduced power for $x_1(\mathbf{s})$ and $x_4(\mathbf{s})$. This trade-off becomes more evident under the nonlinear case. When using $\mathrm{GAM}_L$ for the nonlinear case, the type II error increases more rapidly than for $\mathrm{GAM}_{NL}$ and NW, whereas $\mathrm{GAM}_{NL}$ and NW exhibit higher type I error than $\mathrm{GAM}_L$ as $\theta$ decreases. In general, $\theta = 1$ achieves nominal level control with competitive power. However, when covariates are dependent, the type I/type II trade-off is stronger, and $\theta$ may be tuned according to the user’s objective. For example, around $\theta \approx 0.5$, the random shift methods yield lower type I error rates than the parametric tests—although still above the nominal level—while maintaining comparable or higher power. In the supplementary material, we consider additional scenarios where covariate dependence induces confounding. We observe that the default choice $\theta = 1$ achieves nominal level control while maintaining competitive power as in the dependent case.

\section{Application}
\label{sec: Application}
Chlorophyll-a (Chl-a) is widely used as a proxy for phytoplankton biomass and serves as a key indicator of aquatic ecosystem conditions. Chl-a concentration is influenced by various environmental factors \citep{Kim2022, Liu2025}. We investigate the relationship between Chl-a concentration and environmental covariates, including nitrate concentration ($\mbox{NO}_3$), temperature, pH, and phosphate concentration ($\mbox{PO}_4$), similar to those considered in \cite{Kim2022}.

\begin{figure}[!htbp]
\centering
\vspace{0.2em}
\begin{minipage}[t]{0.49\textwidth}
    \centering
    \includegraphics[width=\textwidth]{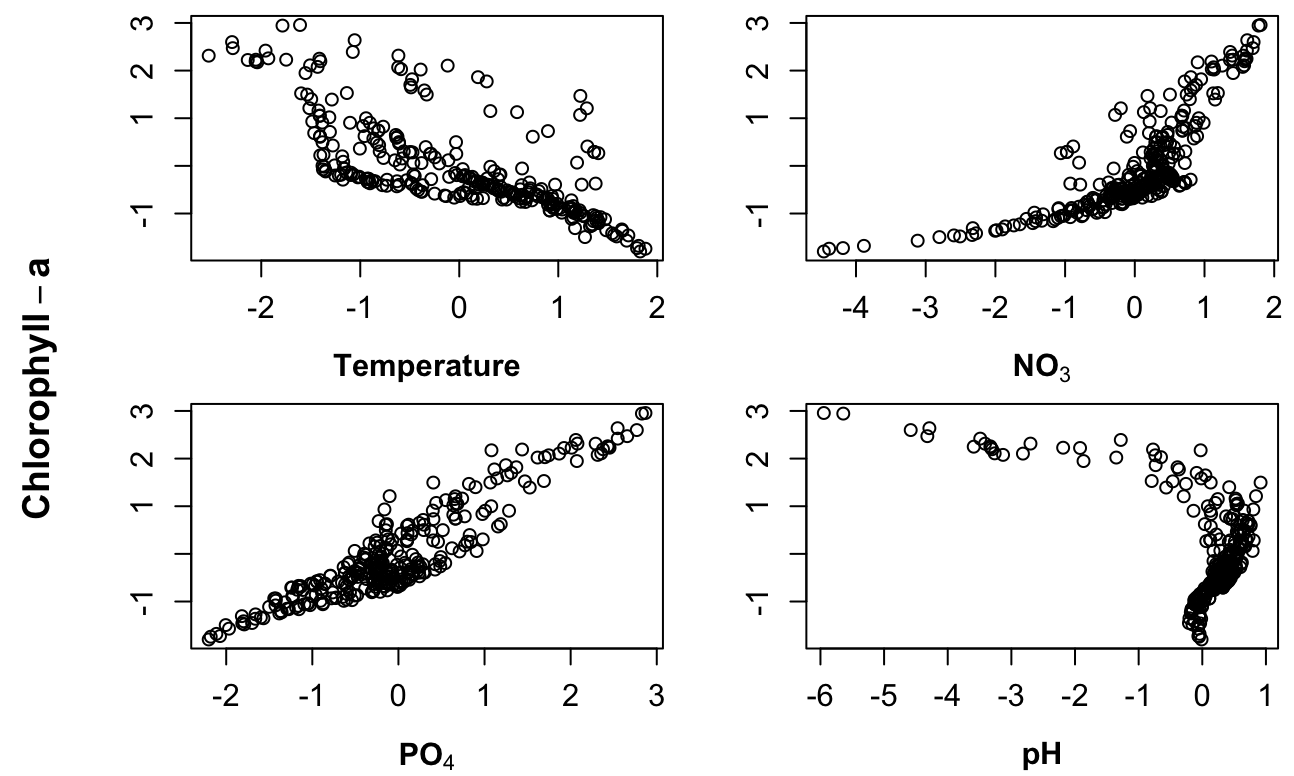}
    \vspace{0.5em}
    {\small (a)}
\end{minipage}
\begin{minipage}[t]{0.49\textwidth}
    \centering
    \includegraphics[width=\textwidth]{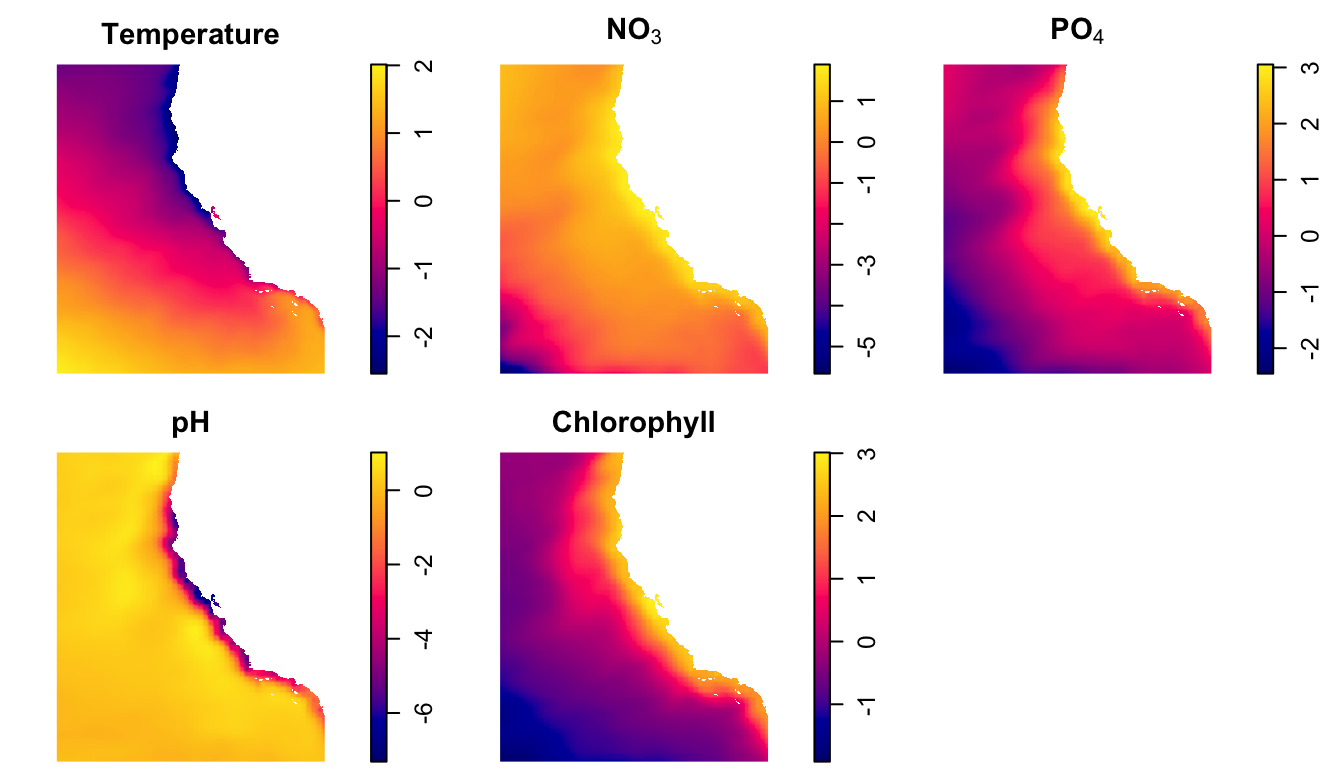}
    \vspace{0.5em}
    {\small (b)}
\end{minipage}
\caption{(a) Scatter plots of the covariates versus the response variable. (b) Spatial maps of the observed variables over the study region.}
\label{fig: realdata_description}
\end{figure}

The data consist of global ocean surface variables measured from 2010 to 2020 on a $0.05^\circ$ spatial grid and were obtained from Bio-ORACLE (\url{https://www.bio-oracle.org/downloads-to-email.php}). The study region spans $30^\circ\mathrm{N}$ to $45^\circ\mathrm{N}$ and $130^\circ\mathrm{W}$ to $117^\circ\mathrm{W}$, covering the coastal ocean near California. A random sample of 300 points from this region was used for analysis. To reduce right-skewness, Chl-a concentration, $\mbox{NO}_3$, and $\mbox{PO}_4$ were log-transformed, and all variables were standardized.

For the nonparametric models (NW, $\mathrm{GAM}_{L}$, and $\mathrm{GAM}_{NL}$), significance testing is performed using the random shift method with variance correction. We consider two test statistics: the distance covariance and the sample covariance. We set $\theta =1$, the default setting. For each test, we generate 999 Monte Carlo samples of the test statistic and perform backward variable selection. For comparison, we also fit an LM, where significance is assessed using a standard $t$-statistic with backward variable selection. In the LM, spatial correlation is modeled using a Mat\'{e}rn covariance function with smoothness parameter 2.5, selected via variogram fitting and AIC.

\begin{table}[htbp]
\centering
\scriptsize
\caption{Variable selection results using the random shift method and the classical method.}
\label{tab:variableselection}
\begin{tabular}{lllc}
\toprule
Method & Test statistic & Model & Significant covariates \\
\midrule
\multirow{6}{*}{Random shift}
& \multirow{3}{*}{Sample covariance}
& $\mathrm{GAM}_{L}$ & $\mathrm{PO}_4$ \\
&  & NW & $\mathrm{PO}_4$ \\
&  & $\mathrm{GAM}_{NL}$ & $\mathrm{PO}_4$ \\

& \multirow{3}{*}{Distance covariance}
& $\mathrm{GAM}_{L}$ & $\mathrm{PO}_4$ \\
&  & NW & $\mathrm{PO}_4$ \\
&  & $\mathrm{GAM}_{NL}$ & $\mathrm{PO}_4$ \\

\midrule
Classical method & $t$-statistic & LM & $\mathrm{PO}_4$, pH \\
\bottomrule
\end{tabular}
\end{table}

Table~\ref{tab:variableselection} shows that the random shift methods identify $\mbox{PO}_4$ as a significant covariate, whereas the classical test selects both $\mbox{PO}_4$ and pH. A possible explanation is the liberality of the classical test when the model is misspecified, for example, in the presence of nonlinear dependence, which is consistent with our studies in Section~\ref{sec: Simulation Study}. Indeed, based on Figure~\ref{fig: realdata_description}, it is difficult to conclude that pH has a significant relationship with Chl-a concentration. In contrast, the relationship between Chl-a concentration and $\mbox{PO}_4$ appears more pronounced. Moreover, although $\mbox{NO}_3$ and temperature also appear to be related to Chl-a concentration, the scatter plots in the supplementary material show clear collinearity among the covariates. In such situations, the effects of $\mbox{NO}_3$ and temperature can be captured by a representative covariate, $\mbox{PO}_4$. The selection of $\mbox{PO}_4$ is aligned with previous studies in coastal marine environments, where phosphate consistently emerged as a significant predictor of Chl-a concentration and dynamics \citep{Liu2025}. This result indicates that the random shift methods are more robust to the confounding case.

\section{Discussion}
\label{sec: Discussion}

We develop fully nonparametric testing procedures for spatial regression. We conduct a Monte Carlo test using test statistics that measure the correlation between the residuals and the covariate of interest. By removing the effects of nuisance covariates, this approach is more robust to collinearity and confounding. The method does not rely on any model assumptions, allowing flexibility. Through extensive numerical studies, we show that our method achieves the nominal rate and outperforms parametric approaches. Theoretical results show that the random shift procedure with variance correction is asymptotically exact in the increasing-domain setting.

Parametric methods \citep{cressie2015statistics, RimalovaEtAl2022} require a correct specification of the models. However, misspecification is likely to occur in spatial settings, where more assumptions are required than in independent data. In such cases, parametric methods tend to become liberal, as we observe in Section~\ref{sec: Simulation Study}. On the other hand, the proposed method provides robust and accurate inference, making it more attractive. We recommend the variance correction because it is asymptotically exact and applicable to irregular domains, whereas the torus correction requires a regular rectangular domain due to the wrap-around mapping along fixed coordinate axes.

As noted, the hyperparameter $\theta$ in \eqref{eq: regress nuisance} controls the extent to which dependence on the covariate of interest is retained in the nuisance covariates. When $\theta=1$, the residuals in \eqref{eq: residuals} remove the signal shared with correlated covariates. Thus, any detected significance is unlikely to be due to confounding, as the confounding effect has been removed. Therefore, our method is more robust to confounding, as observed in the numerical studies. In contrast, when $\theta < 1$, the residuals retain more signal from correlated covariates, which can increase power to a level comparable to parametric methods but also introduce liberality. Therefore, we recommend choosing $\theta=1$, as it achieves the nominal rate and is robust to confounding effects. Furthermore, when covariates are uncorrelated, $\theta=1$ does not compromise power. Developing a data-driven choice of $\theta$ would be an interesting avenue for future research.




\section*{Acknowledgement}
We would like to thank the Editor, the Associate Editor, and the anonymous reviewers for their careful reading and valuable comments. 

\section*{Funding}
This work was supported by the National Research Foundation of Korea (RS-2025-00513129). J. Mateu has been partially funded by a grant PID2022-141555OB-I00 from the Spanish Ministry of Science and Innovation.

\section*{Disclosure statement}\label{disclosure-statement}
The authors report there are no competing interests to declare.

\section*{Data Availability Statement}\label{data-availability-statement}
The source code and data are available at {\url{https://github.com/Whiskeyindia7/NTSR}}. 

\section*{Supplementary Material}
The online supplementary material contains proofs of the main theoretical results, model details, additional simulation studies, and supplementary figures.

\clearpage
\appendix

\begin{center}
\title{\LARGE\bf Supplementary Material for ``Robust Nonparametric Testing Approaches for Spatial Regression''}\\~\\
  
\author{\Large{Kanghyun Wi, Hyoeun Kim, Tom\'a\v s Mrkvi\v cka, Jorge Mateu, and Jaewoo Park}}
\end{center}
\maketitle

The supplementary material contains proofs of the main theoretical results, model details, additional simulation studies, and supplementary figures.

\section{Proofs of the Main Theoretical Results}

\subsection{Proof of Theorem 1}
Let $R_j^n = 1 + \sum_{i=0, i\neq j}^K 1(T_i^n < T_j^n)$ denote the rank of $T_j^n$ among $(T_0^n, \ldots, T_K^n)$ arranged in ascending order. Then $(R_0^n, \ldots, R_K^n)$ is asymptotically a random permutation of $(1, \ldots, K+1)$. Because the test statistics $(T_0^n, \ldots, T_K^n)$ are asymptotically exchangeable, their ranks are also asymptotically exchangeable. Therefore, the limiting marginal distribution of any rank $R_j^n$ is discrete uniform over $\{1, \ldots, K+1\}$. That is, for $j=0, \ldots, K$,
\begin{equation}
    \nonumber
    \lim_{n\rightarrow\infty}\mathbb{P}(R_j^n = k) = \frac{1}{K+1} \quad \text{for } k=1, \ldots, K+1.
\end{equation}

Since $p^n = \frac{1}{K+1} R_0^n$, and $\alpha(K+1)$ is assumed to be an integer, it follows that
\begin{equation}
    \nonumber
    \lim_{n\rightarrow\infty}\mathbb{E}[1(p^n \leq \alpha)] = \lim_{n\rightarrow\infty}\mathbb{P}\{R_0^n \leq \alpha(K+1)\}  = \frac{\alpha(K+1)}{K+1} = \alpha.
\end{equation}

\subsection{Proof of Theorem 2}

Let $\mathbf{c} = (c_0, \ldots, c_K)^\top \in \mathbb{R}^{K+1}$ be an arbitrary non-zero constant vector. By the Cram\'er--Wold device, to establish the joint asymptotic normality of $\mathbf{D}_n^{1/2}\mathbf{T}^n$, it suffices to show that the linear combination $S_n(\mathbf{c}) = \mathbf{c}^\top\mathbf{D}_n^{1/2}\mathbf{T}^n$ converges in distribution to $\mathcal{N}(0, \mathbf{c}^\top \boldsymbol{\Sigma} \mathbf{c})$. Define $S_n(\mathbf{c})$ and $S_n^*(\mathbf{c})$ as
\begin{equation}
    \begin{gathered}
        \nonumber
        S_n(\mathbf{c}) = \sum_{k=0}^K \frac{c_k}{\sqrt{n_k}} \sum_{\mathbf{s}_{i} \in \mathcal{W}_k^n} Y_k(\mathbf{s}_{i}),\quad Y_k(\mathbf{s}) = \Phi(\mathbf{s})\Psi(\mathbf{s} - \mathbf{v}_k), \\
        S_n^*(\mathbf{c}) = \frac{1}{\sqrt{n}} \sum_{\mathbf{s}_{i}\in\mathcal{W}^n} Z_{\mathbf{c}}(\mathbf{s}_{i}), \quad Z_{\mathbf{c}}(\mathbf{s}) = \sum_{k=0}^K c_k Y_k(\mathbf{s}).
    \end{gathered}
\end{equation}

Since $\Phi$ and $\Psi$ are strictly stationary and independent under the null hypothesis, $Z_{\mathbf{c}}(\mathbf{s})$ inherits the zero-mean and strong mixing properties. Under Assumptions~1--4, Theorem~3.2 in \cite{Lahiri} implies that $S^*_n(\mathbf{c}) \overset{d}{\to} \mathcal{N}(0, \sigma^2_\mathbf{c})$, where the asymptotic variance is given by
\begin{equation}
    \begin{aligned}
        \nonumber
        \sigma^2_\mathbf{c} &= \sum_{k=0}^K \sum_{l=0}^K c_k c_l \left\{ C_\Phi(\mathbf{0}) C_\Psi(\mathbf{v}_l - \mathbf{v}_k) + C_2 \int_{\mathbb{R}^2} C_\Phi(\mathbf{h}) C_\Psi(\mathbf{h} - (\mathbf{v}_k - \mathbf{v}_l)) d\mathbf{h} \right\}\\
        &= \sum_{k=0}^K \sum_{l=0}^K c_k c_l {\Sigma}_{kl}.
    \end{aligned}
\end{equation}

Let $\boldsymbol{\Sigma}$ be the $(K+1)\times (K+1)$ matrix whose $(k,l)$-th entry is $\Sigma_{kl}$. We now show that $\boldsymbol{\Sigma}$ is diagonal. For the diagonal elements, the expression simplifies to
\begin{equation}
    \nonumber
    {\Sigma}_{kk} = C_\Phi(\mathbf{0}) C_\Psi(\mathbf{0}) + C_2 \int_{\mathbb{R}^2} C_\Phi(\mathbf{h}) C_\Psi(\mathbf{h}) d\mathbf{h} \equiv \sigma^2, \quad\forall k.
\end{equation}
For the off-diagonal elements ($k \neq l$), Assumption~5 immediately implies that $C_\Psi(\mathbf{v}_l - \mathbf{v}_k) = 0$. Furthermore, note that the integrand in $\Sigma_{kl}$ is nonzero if and only if both $\|\mathbf{h}\| \leq R$ and $\|\mathbf{h} - (\mathbf{v}_k - \mathbf{v}_l)\| \leq R$ by Assumption~5. By the triangle inequality, the existence of such $\mathbf{h}$ would require
\begin{equation}
    \nonumber
    \|\mathbf{v}_k - \mathbf{v}_l\| \leq \|\mathbf{h}\| + \|\mathbf{h} - (\mathbf{v}_k - \mathbf{v}_l)\| \leq 2R.
\end{equation}
However, this contradicts the condition in Assumption~6. Thus, no such $\mathbf{h}$ exists; consequently, $\Sigma_{kl} = 0$ for all $k \neq l$. Therefore, we have
\begin{equation}
    \nonumber
    S_n^*(\mathbf{c}) \overset{d}{\to} \mathcal{N}(0, \mathbf{c}^\top (\sigma^2\mathbf{I}) \mathbf{c}) \quad \text{as } n\to\infty.
\end{equation}

To complete the proof, we show that $S_n(\mathbf{c}) - S_n^*(\mathbf{c}) \overset{p}{\to} 0$. Define the non-overlap region $B^n_k = \mathcal{W}^n \setminus \mathcal{W}^n_k$. By Assumption~1, the Lebesgue measure $|B^n_k|$ is of order $\mathcal{O}(\lambda_n)$ for any finite $\mathbf{v}_k$, so $|\lambda_n^{-1} B^n_k| = \mathcal{O}(1/\lambda_n)$. Since the density $p(\mathbf{u})$ is continuous on the closure of $\mathcal{W}^0$, it is bounded above by some constant $M_{\mathbf{u}} < \infty$. Therefore, the probability that a point falls into $B^n_k$ is bounded by $M_{\mathbf{u}} |\lambda_n^{-1} B^n_k| = \mathcal{O}(1/\lambda_n)$. Because the points are drawn independently, the number of points in 
$B^n_k$, denoted by $b_k$, follows a binomial distribution with expectation
\begin{equation}
    \nonumber
    \mathbb{E}[b_k]\leq nM_{\mathbf{u}} |\lambda_n^{-1} B^n_k| = \mathcal{O}(\lambda_n^2)\times\mathcal{O}(1/\lambda_n) = \mathcal{O}(\lambda_n).
\end{equation}
By Markov's inequality, since $n/\lambda_n^2 \to C_1 \in (0,\infty)$, we have
\begin{equation}\label{eq: boundary_ratio}
    \frac{b_k}{n} = \mathcal{O}_p\left(\frac{1}{\lambda_n}\right) = o_p(1), \quad \text{and} \quad \frac{n_k}{n} = 1 - \frac{b_k}{n} \overset{p}{\to} 1.
\end{equation} 

Decompose the difference $S_n(\mathbf{c}) - S^*_n(\mathbf{c})$ into two components as follows:
\begin{equation}\label{eq: decomp}
    S_n(\mathbf{c}) - S_n^*(\mathbf{c}) = \sum_{k=0}^K c_k \underbrace{ \left( \sqrt{\frac{n}{n_k}} - 1 \right) \left( \frac{1}{\sqrt{n}} \sum_{\mathbf{s}_{i} \in \mathcal{W}_k^n} Y_k(\mathbf{s}_{i}) \right) }_{ \text{(A)} } - \sum_{k=0}^K c_k \underbrace{ \frac{1}{\sqrt{n}} \sum_{\mathbf{s}_{i} \in B^n_k} Y_k(\mathbf{s}_{i}) }_{ \text{(B)} }.
\end{equation}

For term (A), equation \eqref{eq: boundary_ratio} implies that 
$(\sqrt{n/n_k} - 1) = o_p(1)$. Since the remaining summation term is $\mathcal{O}_p(1)$ by the same central limit theorem in \cite{Lahiri}, term (A) converges to zero in probability.

For term (B), let $I_i=1(\mathbf{s}_{i} \in B_k^n)$ be the indicator variable that equals one if the $i$-th point falls into the non-overlap region, and zero otherwise. By the law of total variance, the variance of term (B) is given by
\begin{equation}
   	\begin{aligned}\label{eq: term B variance}
   		\mathrm{Var}\left( \frac{1}{\sqrt{n}} \sum_{\mathbf{s}_{i}\in B^n_k}  Y_k(\mathbf{s}_{i}) \right) &= \frac{1}{n} \mathbb E\left[ \mathrm{Var}\left(  \sum_{i=1}^n I_i Y_k(\mathbf{s}_{i})\mid \mathbf u_{1},\dots, \mathbf u_{n} \right) \right]\\
   		&= \frac{1}{n}\sum_{i=1}^n\mathbb E\left[I_i \mathrm{Var}(Y_k(\mathbf 0 ))  \right] + \frac{1}{n} \sum_{i\neq j}  \mathbb E\left[ I_i I_j \mathrm{Cov} (Y_k(\mathbf s_i), Y_k(\mathbf s_j))\right]\\
   	\end{aligned}
\end{equation}
The first summation term in \eqref{eq: term B variance} can be written as
\begin{align*}
	\frac{1}{n}\sum_{i=1}^n\mathbb E\left[I_i \mathrm{Var}(Y_k(\mathbf 0 ))\right] &= \mathrm{Var}(Y_k(\mathbf 0 ))\mathbb{P}(\lambda_n \mathbf{u}_1 \in B_k^n) \\
	&\leq \mathrm{Var}(Y_k(\mathbf 0 ))M_\mathbf u|\lambda_n^{-1}B^n_k| = \mathcal O \left(\frac{1}{\lambda_n}\right),
\end{align*}
which converges to zero as $n \to \infty$.

For the second term, we have
\begin{equation}
    \begin{aligned}
    	\nonumber
    	\mathbb{E}[I_i I_j \mathrm{Cov} (Y_k(\mathbf{s}_{i}), Y_k(\mathbf{s}_{j}))] &= \iint_{\lambda_n\mathbf{u}, \lambda_n\mathbf{t} \in B_k^n} \mathrm{Cov}(Y_k(\lambda_n\mathbf{u}), Y_k(\lambda_n\mathbf{t})) p(\mathbf{u}) p(\mathbf{t}) d\mathbf{u} d\mathbf{t}\\
        &\leq \frac{M_\mathbf{u}^2}{\lambda_n^2} \int_{\lambda_n^{-1}B_k^n} \left(  \int_{\mathbb{R}^2} |\mathrm{Cov}(Y_k(\mathbf 0 ), Y_k(\mathbf h))| d\mathbf{h} \right) d\mathbf{t}\\
        &= \frac{M_\mathbf{u}^2}{\lambda_n^2} \int_{\lambda_n^{-1}B_k^n} \left(  \int_{\mathbb{R}^2} |C_{\Phi}(\mathbf h) C_{\Psi}(\mathbf h)| d\mathbf{h} \right) d\mathbf{t}\\
        &= \frac{M_\mathbf{u}^2}{\lambda_n^2} \int_{\lambda_n^{-1}B_k^n} \bigg( \int_{\|\mathbf h\|\leq R} |C_{\Phi}(\mathbf h) C_{\Psi}(\mathbf h)| d\mathbf{h}+ \int_{\|\mathbf h\| > R} |C_{\Phi}(\mathbf h) C_{\Psi}(\mathbf h)| d\mathbf{h}\bigg) d\mathbf{t}.
    \end{aligned}
\end{equation}
The first inner integration term is bounded by a constant such that
\begin{equation}
    \nonumber
    \int_{\|\mathbf h\|\leq R} |C_{\Phi}(\mathbf h) C_{\Psi}(\mathbf h)| d\mathbf{h} \leq \pi R^2 C_{\Phi}(\mathbf 0) C_{\Psi}(\mathbf 0) \equiv M_C < \infty,
\end{equation}
and the second integration term is identically zero by Assumption~5. Consequently, the second summation term in \eqref{eq: term B variance} is bounded as
\begin{equation}
    \nonumber
    \frac{1}{n} \sum_{i\neq j}  \mathbb E\left[ I_i I_j \mathrm{Cov} (Y_k(\mathbf s_i), Y_k(\mathbf s_j))\right] \leq \frac{n(n-1)}{n}\times \frac{M_\mathbf{u}^2 M_C}{\lambda_n^2} |\lambda_n^{-1}B_k^n| = \mathcal{O}\left(\lambda_n^2\right)\times\mathcal{O}\left(\frac{1}{\lambda_n^3}\right),
\end{equation}
which converges to zero as $n\to\infty$.

Because both the diagonal and off-diagonal components vanish, 
term (B) converges to zero in mean square and hence in probability. 
Since terms (A) and (B) in equation \eqref{eq: decomp} both converge to zero in probability, $S_n(\mathbf{c}) - S_n^*(\mathbf{c}) \overset{p}{\to} 0$. By Slutsky's theorem, $S_n(\mathbf{c})$ has the same limiting distribution as $S_n^*(\mathbf{c})$. Because this holds for any non-zero constant vector $\mathbf{c}$, the Cram\'er--Wold device yields the joint asymptotic normality:
\begin{equation}
    \nonumber
    \mathbf{D}_n^{1/2}{\mathbf{T}}^n \overset{d}{\to} \mathcal{N}_{K+1}(\mathbf{0}, \sigma^2\mathbf{I}).
\end{equation}
This completes the proof.

\clearpage

\section{Model Details}
\subsection{Details of the Fitted Model}

\begin{enumerate}
\item Spatial linear model (LM)
\[
y(\mathbf{s})= \beta_0+x_1(\mathbf{s})\beta_1 + \epsilon(\mathbf{s}),
\]
where $\epsilon(\mathbf{s})$ follows a stationary Gaussian process with a squared exponential covariance function $C(\mathbf{s}_i,\mathbf{s}_j)=\sigma^2 \exp(-\|\mathbf{s}_i-\mathbf{s}_j\|^2/\phi^2) + \tau^2$.
\item Nadaraya-Watson kernel estimator (NW)
\[
y(\mathbf{s})= f_1(x_1(\mathbf{s}))+ \epsilon(\mathbf{s}),
\]
where $f_1(x_1(\mathbf{s}))$ is estimated using the following Nadaraya-Watson kernel estimator:
\[
    \hat f(x_1(\mathbf{s})) = \frac{\sum_{i=1}^n K_h(x_1(\mathbf{s}) - x_1(\mathbf{s}_i))y(\mathbf{s}_i)}{\sum_{i=1}^n K_h(x_1(\mathbf{s}) - x_1(\mathbf{s}_i))}.
\]
Here, $K_h(u) = K(u / h)$ is a kernel function, and we use a second-order Epanechnikov kernel function defined as $K(u/h)=3(1-(u/h)^2/5)/(4 \sqrt{5})$ if $(u/h)^2 < 5$, 0 otherwise with bandwidth parameter $h>0$, originally proposed by \cite{Epanechnikov1969}. The bandwidth is selected using the least squares cross-validation method by \citet{Racine2004} and \citet{Li2004}, as implemented in the \texttt{npregbw} function of the \texttt{np} package.
\item GAM with a linear mean structure ($\mbox{GAM}_{L}$)
\[
y(\mathbf{s})= \beta_0+x_1(\mathbf{s})\beta_1 + \psi(\mathbf{s})+ \epsilon(\mathbf{s}),
\]
where $\psi(\mathbf{s})$ is modeled using thin plate regression splines of order $m$ \citep{Wood2003}. Specifically, $\psi(\mathbf{s})=\sum_{i=1}^n\delta_i \eta_{m,2}(\|\mathbf{s}-\mathbf{s}_i\|)+\sum_{j=1}^M\alpha_j\phi_j(\mathbf{s})$ where $\bm \delta$ and $\bm \alpha$ are unknown parameter vectors subject to the constraint $\mathbf{T}^{\top} \bm \delta= \mathbf{0}$, with $\mathbf{T}_{ij}=\phi_j(\mathbf{s}_i)$. A total of $M=\binom {m+d-1}{d}$ basis functions $\phi_i$ are used, forming a linearly independent basis for the space of polynomials in $\mathbb{R}^d$ of degree less than $m$. The function $\eta_{m, 2}$ denotes the radial basis function of order $m$ in dimension $d=2$ defined as follows,
\begin{equation*}
    \eta_{m,d}(r) =
    \begin{cases}
     \frac{(-1)^{m + 1 + d/2}}{2^{2m-1} \pi^{d/2} (m-1)! (m - d/2)!} \, r^{2m - d} \log(r) & \text{if } d \text{ is even} \\ 
     \frac{\Gamma(d/2 - m)}{2^{2m} \pi^{d/2} (m-1)!} \, r^{2m - d} & \text{if } d \text{ is odd}
     \end{cases}
\end{equation*}
\item GAM with a nonlinear mean structure ($\mbox{GAM}_{NL}$)
\[
y(\mathbf{s})= \beta_0+f_1(x_1(\mathbf{s})) + \psi(\mathbf{s})+ \epsilon(\mathbf{s}).
\]
Both $\psi(\mathbf{s})$ and $f_1(x_1(\mathbf{s}))$ are modeled using thin plate regression splines \citep{Wood2003}. Specifically, $f_1(x_1(\mathbf{s}))=\sum_{i=1}^{n}\delta^{(1)}_i \eta_{m,1}(\|x_1(\mathbf{s})-x_1(\mathbf{s}_i)\|)+\sum_{j=1}^{M_1}\alpha^{(1)}_j\phi^{(1)}_j(x_1(s))$ and $\psi(\mathbf{s})=\sum_{i=1}^{n}\delta^{(2)}_i \eta_{m,2}(\|\mathbf{s}-\mathbf{s}_i\|)+\sum_{j=1}^{M_2}\alpha^{(2)}_j\phi^{(2)}_j(\mathbf{s})$.
\end{enumerate}

\subsection{Details of the Random Error Structure}

\begin{enumerate}

\item Stationary Gaussian process with unit variance (SE1)

$\epsilon(\mathbf{s})$ follows a stationary Gaussian process with squared exponential covariance
\[
C(\mathbf{s}_i, \mathbf{s}_j)
= \sigma^2 \exp\!\left(-\frac{\|\mathbf{s}_i-\mathbf{s}_j\|^2}{\phi^2}\right) + \tau^2,
\]
where $\sigma^2=1$, $\phi=0.2$, and $\tau^2=1$.

\item Stationary Gaussian process with variance 4 (SE4)

Same as SE1, but with variance $\sigma^2=4$.

\item Exponential Gaussian process (E1)

$\epsilon(\mathbf{s})$ follows a stationary Gaussian process with exponential covariance
\[
C(\mathbf{s}_i, \mathbf{s}_j)
= \sigma^2 \exp\!\left(-\frac{\|\mathbf{s}_i-\mathbf{s}_j\|}{\phi}\right) + \tau^2,
\]
where $\sigma^2=4$, $\phi=0.2$, and $\tau^2=1$.

\item Negative-dependence Gaussian process (N)

To induce negative spatial dependence, the window $\mathcal{W}$ is partitioned into
$16$ disjoint subwindows
\[
\mathcal{W}_{p,q} = \big[(p-1)/4,\, p/4\big] \times \big[(q-1)/4,\, q/4\big],
\quad p,q=1,\ldots,4,
\]
with $\mathcal{W}=\bigsqcup_{p,q}\mathcal{W}_{p,q}$.
We define
\[
\epsilon(\mathbf{s}) = w(\mathbf{s}) + 0.5\,(-1)^{p+q}, \qquad \mathbf{s}\in\mathcal{W}_{p,q},
\]
where $w(\mathbf{s})$ is a Gaussian process with squared exponential covariance as in SE1 ($\sigma^2=1$, $\phi=0.2$, $\tau^2=1$). This construction yields a checkerboard-like pattern of negative spatial dependence.

\item Log-Gaussian process (LN)

$\epsilon(\mathbf{s})$ follows a log-Gaussian process such that
$\log \epsilon(\mathbf{s})$ is a Gaussian process with squared exponential covariance as in SE1, with parameters $\sigma^2=1$, $\phi=0.2$, and $\tau^2=1$.

\item Nonstationary Gaussian process (NS)

We simulate a nonstationary Gaussian process using the kernel convolution-based method \citep{Mark2017} in which local parameters at any location \( \mathbf{s} \) are computed as a weighted average of the parameters at fixed mixture component locations \( \{\mathbf{b}_k\}_{k=1}^{K} \). In our simulation, we set \( K=4 \) with $\mathbf{b}_1 = (0.25, 0.25), \mathbf{b}_2 = (0.75, 0.25), \mathbf{b}_3 = (0.25, 0.75),\mathbf{b}_4 = (0.75, 0.75)$. At each \( \mathbf{b}_k \), local anisotropy parameters—eigenvalues \( \lambda_1(\mathbf{s}), \lambda_2(\mathbf{s}) \) and rotation angle \( \eta(\mathbf{s}) \)—are specified via generalized linear models:
\[
\begin{aligned}
\lambda_1(\mathbf{s}) &= \exp\{\beta_0^{\lambda_1} + \beta_1^{\lambda_1}s_1 + \beta_2^{\lambda_1}s_2\}, \\
\lambda_2(\mathbf{s}) &= \exp\{\beta_0^{\lambda_2} + \beta_1^{\lambda_2}s_1 + \beta_2^{\lambda_2}s_2\}, \\
\eta(\mathbf{s}) &= \frac{\pi}{2} \cdot \text{logit}^{-1}(\beta_0^{\eta} + \beta_1^{\eta}s_1 + \beta_2^{\eta}s_2).
\end{aligned}
\]
We use the following coefficient values: \( \bm{\beta}_{\lambda_1} = (-1.3, 0.5, -0.6) \), \( \bm{\beta}_{\lambda_2} = (-1.4, -0.1, 0.2) \), and \( \bm{\beta}_{\eta} = (0, -0.15, 0.15) \). Then the resulting parameters at each mixture component are computed as $(\lambda_{1,1}, \lambda_{2,1}, \eta_1) = (0.2658, 0.2528, 0.7854), (\lambda_{1,2}, \lambda_{2,2}, \eta_2) = (0.3413, 0.2405, 0.756)$, $(\lambda_{1,3}, \lambda_{2,3}, \eta_3) = (0.1969, 0.2794, 0.8148), (\lambda_{1,4}, \lambda_{2,4}, \eta_4) = (0.2528, 0.2658, 0.7854)$. 

The kernel matrix \( \mathbf{\Sigma}_k \) at each \( \mathbf{b}_k \) is constructed using spectral decomposition:
\[
\mathbf{\Sigma}_k =
\begin{bmatrix}
\cos(\eta_k) & -\sin(\eta_k) \\
\sin(\eta_k) & \cos(\eta_k)
\end{bmatrix}
\begin{bmatrix}
\lambda_{1,k} & 0 \\
0 & \lambda_{2,k}
\end{bmatrix}
\begin{bmatrix}
\cos(\eta_k) & \sin(\eta_k) \\
-\sin(\eta_k) & \cos(\eta_k)
\end{bmatrix}.
\]
The spatially-varying kernel matrix at any location \( \mathbf{s} \) is computed as
\[
\mathbf{\Sigma}(\mathbf{s}) = \sum_{k=1}^{4} w_k(\mathbf{s}) \mathbf{\Sigma}_k, \quad
w_k(\mathbf{s}) \propto \exp\left(-\frac{\|\mathbf{s} - \mathbf{b}_k\|^2}{2\lambda_w}\right), \quad \sum_{k=1}^4 w_k(\mathbf{s}) = 1.
\]
Following  \cite{Mark2017}, the tuning parameter \( \lambda_w \) is computed as $\lambda_w = \left(0.5 \cdot \|\mathbf{b}_1 - \mathbf{b}_2\|\right)^2 = 0.0625$. Finally, the full covariance function is 
\[
C(\mathbf{s}_i,\mathbf{s}_j) = \sigma^2 \rho(\mathbf{s}_i,\mathbf{s}_j) g\left(\sqrt{Q(\mathbf{s}_i, \mathbf{s}_j)}\right),
\]
where
\[
\rho(\mathbf{s}_i,\mathbf{s}_j) = \frac{|\mathbf{\Sigma}(\mathbf{s}_i)|^{1/4} |\mathbf{\Sigma}(\mathbf{s}_j)|^{1/4}}{\left|\frac{\mathbf{\Sigma}(\mathbf{s}_i) + \mathbf{\Sigma}(\mathbf{s}_j)}{2}\right|^{1/2}},
\quad
Q(\mathbf{s}_i, \mathbf{s}_j) = (\mathbf{s}_i - \mathbf{s}_j)^T \left(\frac{\mathbf{\Sigma}(\mathbf{s}_i) + \mathbf{\Sigma}(\mathbf{s}_j)}{2}\right)^{-1} (\mathbf{s}_i - \mathbf{s}_j),
\]
and \( g(\cdot) \) is a Mat\'ern correlation function with smoothness $\kappa$. In our study, we set \( \sigma^2 = 1 \) and \( \kappa = 0.5 \) and simulate a nonstationary process through the \texttt{NSconvo\_sim} function in the \texttt{convoSPAT} package. 
\end{enumerate}

\clearpage
\section{Additional Experiments}

Another choice for the test statistic is a correlation coefficient. Following \cite{Dvorak2024}, we investigate using Kendall's correlation coefficient due to its nonparametric nature. The empirical estimator of Kendall's $\tau$ is defined as
\begin{equation}
	\nonumber
    \widehat \tau = \frac{1}{n(n-1)} \sum_{i \not= j}\operatorname{sgn}\big(x_{d+1}(\mathbf{s}_i) - x_{d+1}(\mathbf{s}_j)\big)  \operatorname{sgn}\big(e(\mathbf{s}_i) - e(\mathbf{s}_j)\big),
\end{equation}
where $\operatorname{sgn}(\cdot)$ denotes the sign function. Since \cite{Vaart1998} showed that the asymptotic order of $\widehat \tau$ follows $1/n$ when we have $n$ observations, we use $1/n_k$ as an estimate of $\operatorname{var}(T_k)$ for $\widehat{\tau}$, where $n_k$ denotes the number of observations in the intersection window $\mathcal{W}_k$.

\subsection{Simulation Results with Kendall’s Correlation and Torus Correction}

\begin{figure}[htbp]
\centerline{\includegraphics[width=\textwidth]{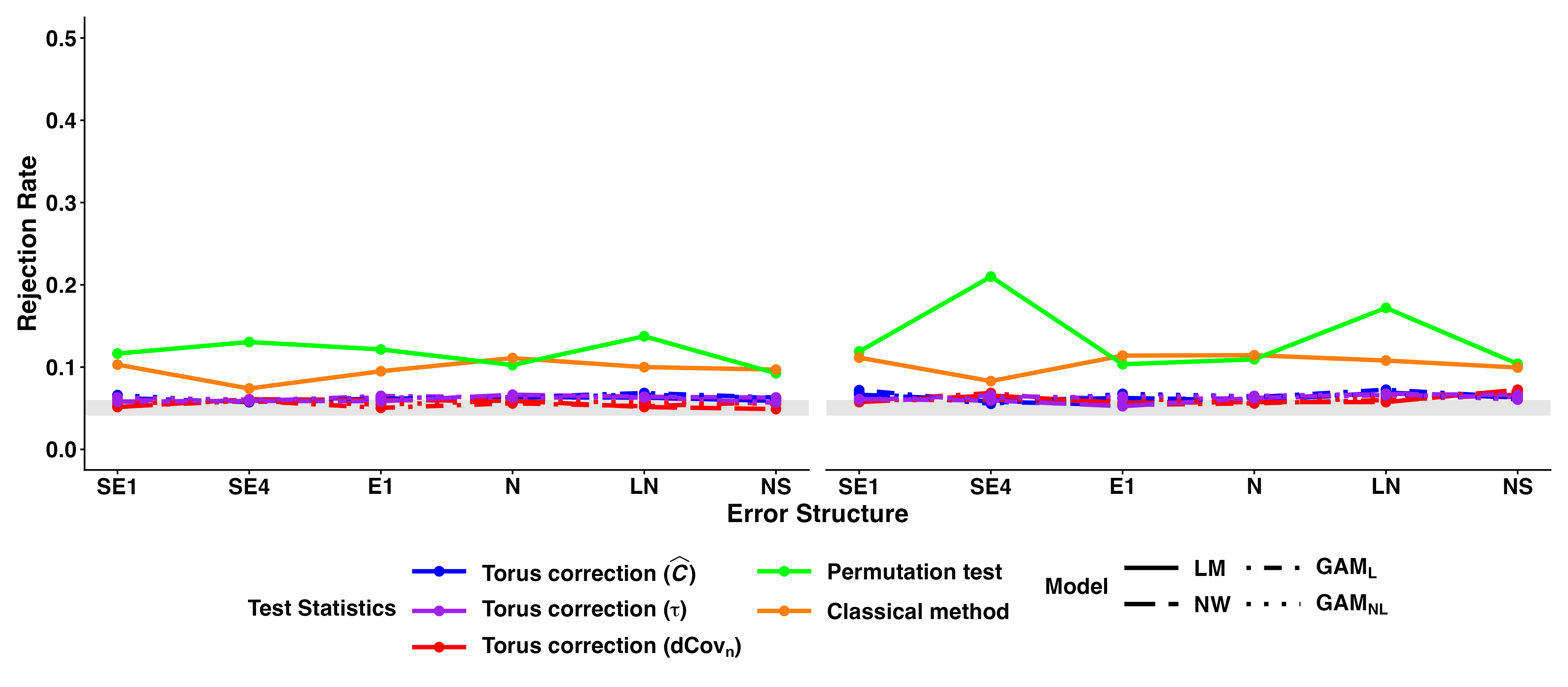}}
\caption{Empirical rejection rates under the linear model simulation (left) and nonlinear model simulation (right) using torus correction. The shaded horizontal band indicates the 95\% binomial confidence interval for the rejection rate at $\alpha=0.05$, given by [0.041, 0.060].}
\label{torus_sig}
\end{figure}

We conduct simulations under the same simulation design as in Section~5.1, adding Kendall's correlation coefficient as a test statistic. For random shift methods, torus correction is applied. Figure~\ref{torus_sig} shows that the results using Kendall’s correlation coefficient and torus correction are quantitatively similar to those based on variance correction in the main manuscript. In both linear and nonlinear models, random shift methods yield rejection rates close to the nominal level in all scenarios, regardless of the choice of test statistic. This indicates that random shift methods are robust and provide proper control of type I error. In contrast, the classical method and the permutation test exhibit more liberality across all scenarios compared to the random shift methods. 


\subsection{Simulation Results with Varying Sample Size}

We perform simulations with the same design as in Section~5.1, but with varying sample sizes. Specifically, we consider three cases: (1) $\mathcal W=[0,0.5]\times[0,0.5]$ with 25 observations, (2) $\mathcal W=[0,2]\times[0,2]$ with 400 observations, and (3) $\mathcal W=[0,3]\times[0,3]$ with 900 observations. As both the window size and the number of observations increase, the permutation test becomes numerically unstable and liberal, whereas the classical method exhibits reduced liberality, with empirical rejection rates approaching the nominal confidence interval. In contrast, the difference between the two random shift methods becomes negligible for larger windows and sample sizes.

\begin{enumerate}
\item {$\mathcal W=[0 ,0.5]\times [0 ,0.5]$ with 25 observations}

Figure~\ref{25_sig} summarizes the results under both linear and nonlinear trend models. Compared with the unit-square window (100 observations) case, the classical method is the most liberal across testing procedures in both settings. The permutation test also exhibits liberal behavior, although its degree of liberality is smaller than that observed in the unit square case. Under the linear setting, the empirical rejection rates of the permutation test for the SE1, E1, N, and NS scenarios fall within the nominal confidence interval, whereas under the nonlinear setting, those for the E1 and NS scenarios lie near the boundary of the nominal level. Overall, the random shift methods demonstrate greater robustness than the competing approaches, with the torus correction consistently being more liberal than the variance correction.

\begin{figure}[htbp]
\centerline{\includegraphics[width=\textwidth]{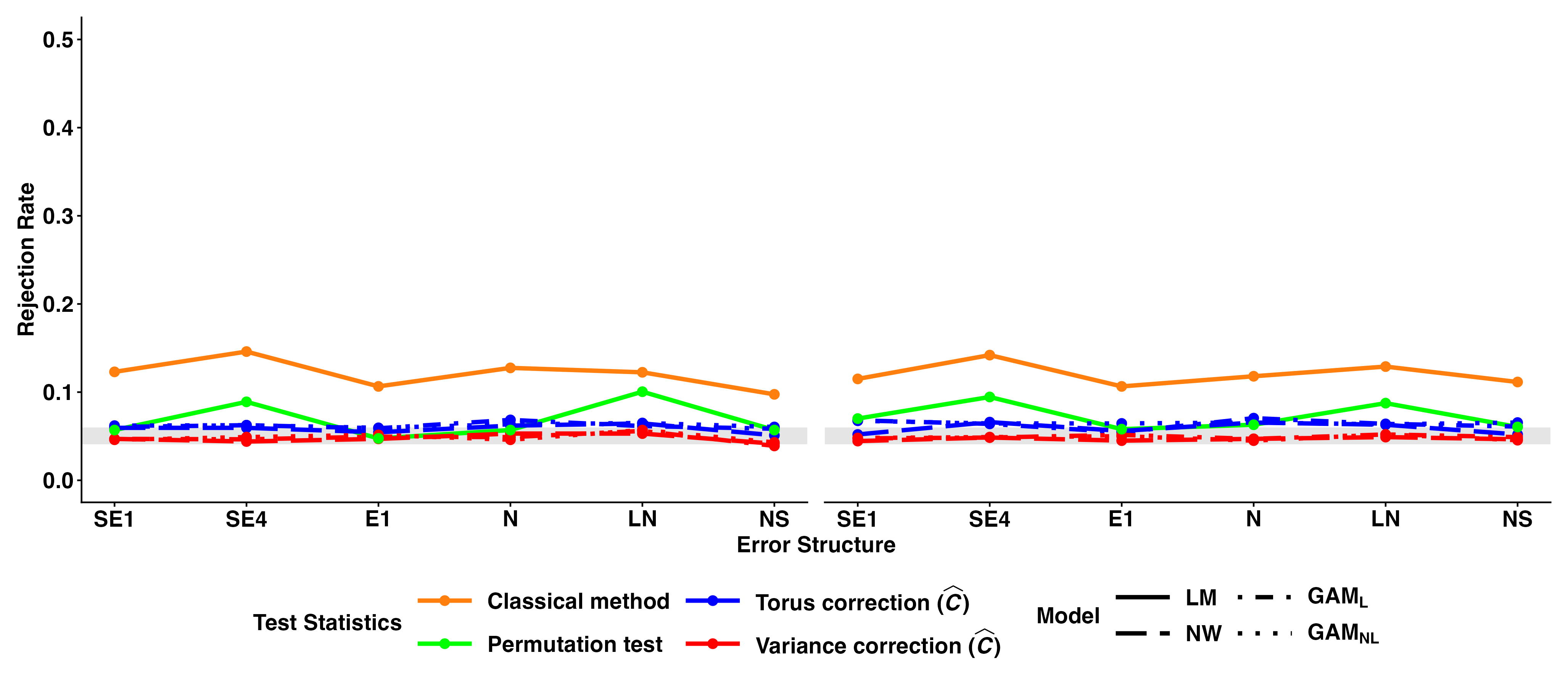}}
\caption{Empirical rejection rates under the linear (left) and nonlinear (right) simulation models with 25 observations. The shaded horizontal band indicates the 95\% binomial confidence interval for the rejection rate at $\alpha=0.05$, given by [0.041, 0.060].}
\label{25_sig}
\end{figure}

\item {$\mathcal W=[0 ,2]\times[0,2]$ with 400 observations}

Figure~\ref{400_sig} shows that the permutation test is the most liberal across both settings, exhibiting substantial liberality in all scenarios. In contrast, the classical method becomes less liberal, with rejection rates approaching the nominal level. The two correction variants of the random shift method yield comparable results, with empirical rejection rates mostly lying within or near the boundary of the nominal confidence interval.

\begin{figure}[htbp]
\centerline{\includegraphics[width=\textwidth]{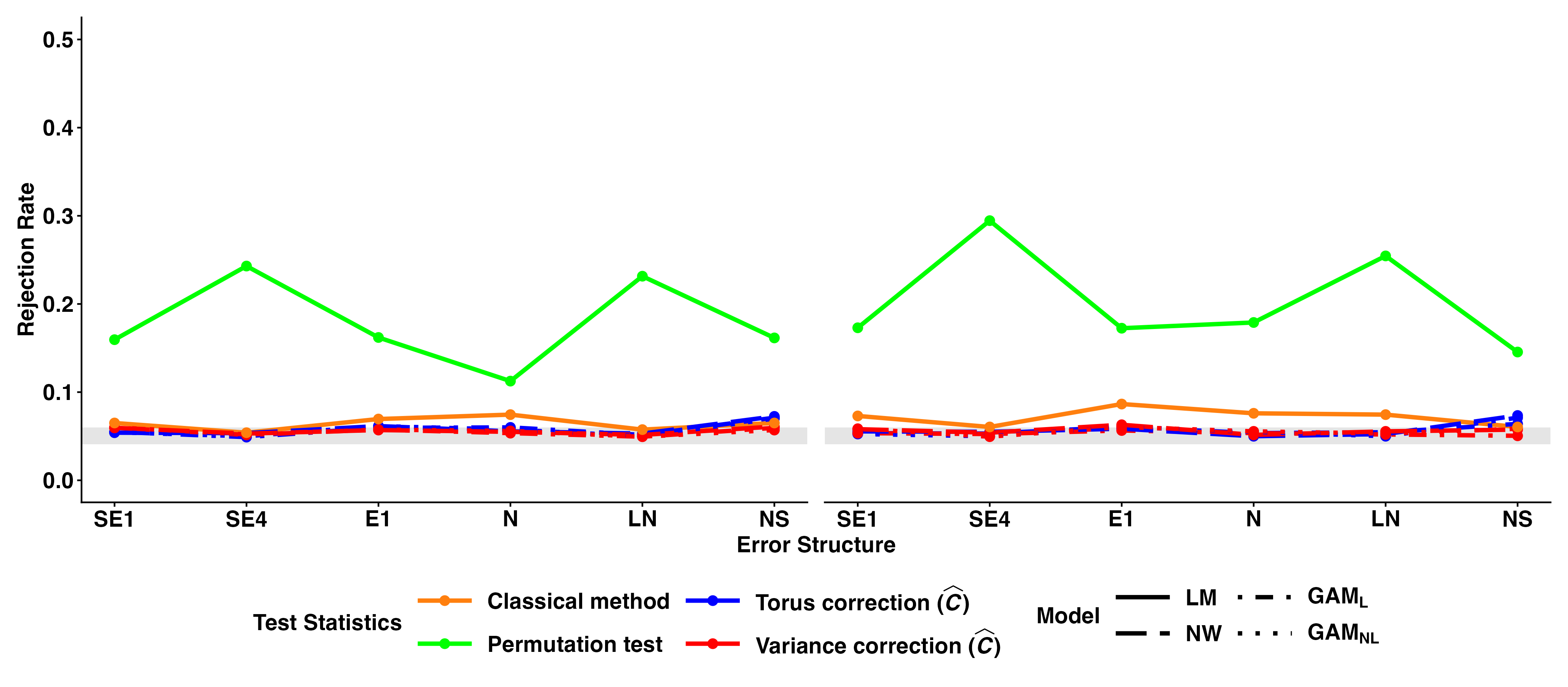}}
\caption{Empirical rejection rates under the linear (left) and nonlinear (right) simulation models with 400 observations. The shaded horizontal band indicates the 95\% binomial confidence interval for the rejection rate at $\alpha=0.05$, given by [0.041, 0.060].}
\label{400_sig}
\end{figure}

\item {$\mathcal W=[0 ,3]\times[0,3]$ with 900 observations}

Figure~\ref{900_sig} shows that the permutation test exhibits the greatest instability and substantial liberality relative to settings with smaller windows and fewer observations. In contrast, the classical method produces more stable and less liberal results in both settings, comparable to those obtained with the random shift methods. The random shift methods with torus and variance correction yield similar results across all scenarios.

\begin{figure}[htbp]
\centerline{\includegraphics[width=\textwidth]{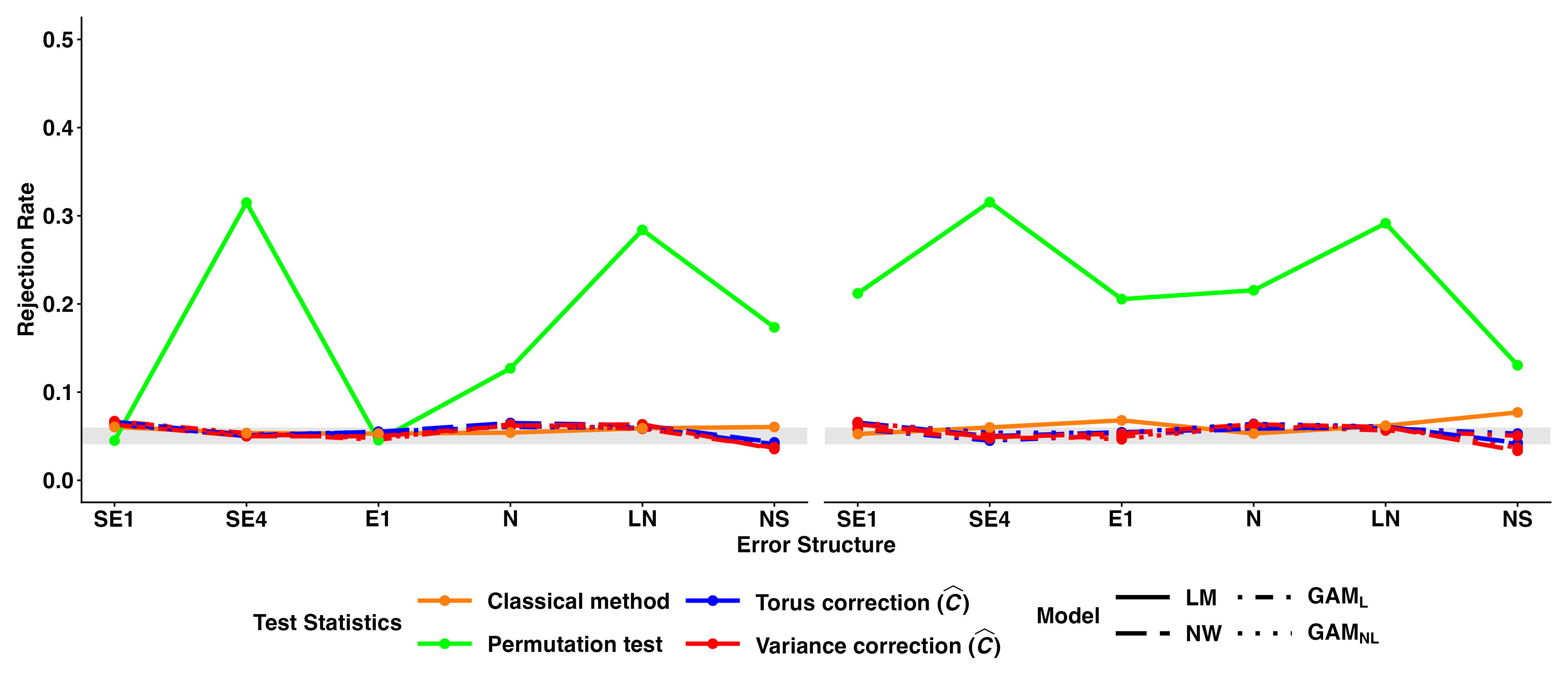}}
\caption{Empirical rejection rates under the linear (left) and nonlinear (right) simulation models with 900 observations. The shaded horizontal band indicates the 95\% binomial confidence interval for the rejection rate at $\alpha=0.05$, given by [0.041, 0.060].}
\label{900_sig}
\end{figure}

\end{enumerate}

\subsection{Power Analysis}

We conduct a power analysis under the same simulation setting as in Section~5.1. Figure~\ref{variance(cov, dcov)_pow} illustrates the empirical power under linear and nonlinear mean trends. Under the linear trend, the classical method achieves the highest power, while the random shift methods exhibit slightly lower power than the classical and permutation tests, consistent with their conservative behavior. A similar pattern is observed under the nonlinear trend: when using the sample covariance with variance correction, the random shift method consistently yields lower power than the classical and permutation tests. In contrast, when distance covariance is used as the test statistic, the random shift method generally attains higher power than the competing approaches.

\begin{figure}[htbp]
\centerline{\includegraphics[width=\textwidth]{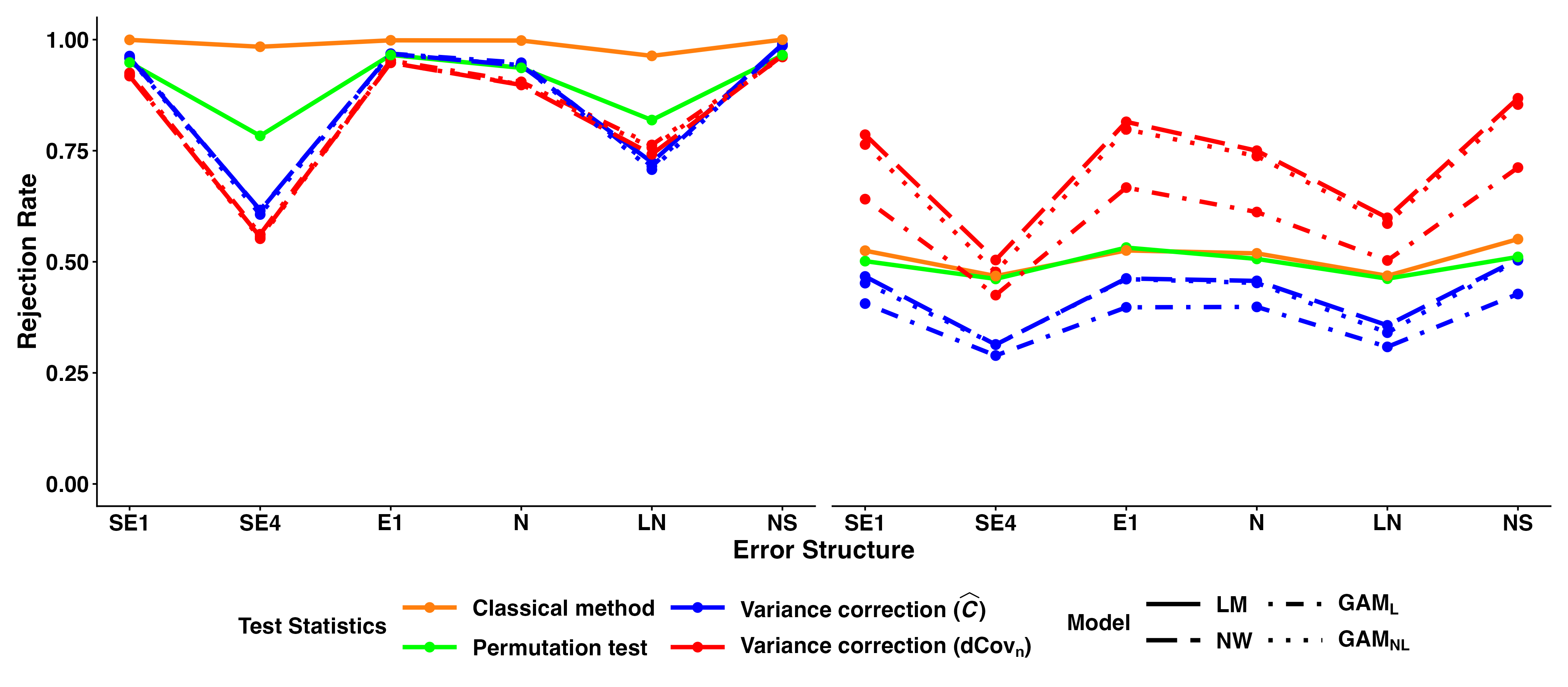}}
\caption{Empirical rejection rates under the linear model simulation (left) and nonlinear model simulation (right).}
\label{variance(cov, dcov)_pow}
\end{figure}

\subsection{Multiple Dependent Covariates with Confounding}

We conduct simulations with the same design as in Section 5.3, except that $x_4(\mathbf{s})$ does not contribute to the generation of $y(\mathbf{s})$, thereby inducing confounding. Therefore, we evaluate type I error control for $x_3(\mathbf{s})$ and $x_4(\mathbf{s})$, which are not included in the data-generating model, and empirical power for $x_1(\mathbf{s})$ and $x_2(\mathbf{s})$.

\begin{figure}[htbp]
\centerline{\includegraphics[width=\textwidth]{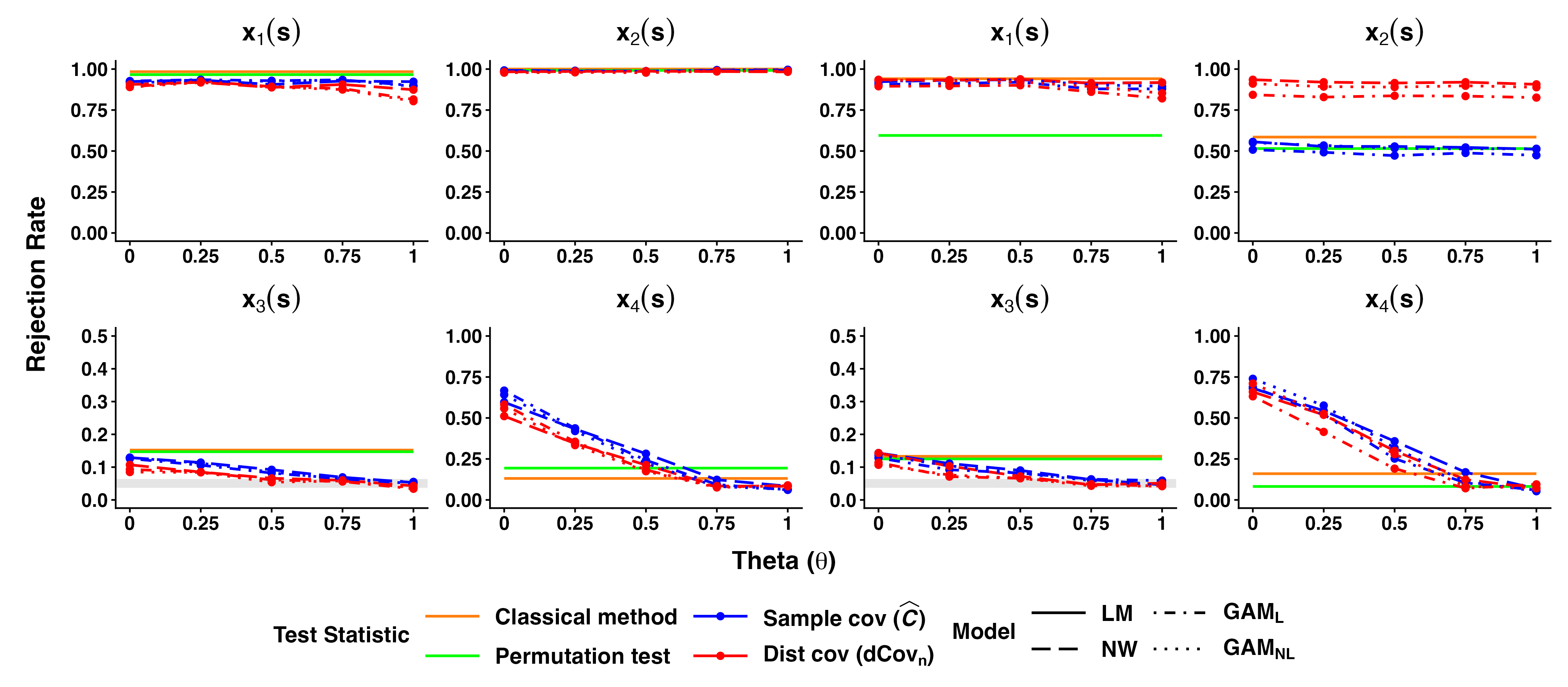}}
\caption{Empirical rejection rates under the linear model (left) and the nonlinear model (right). The shaded horizontal band indicates the 95\% binomial confidence interval for the rejection rate at $\alpha=0.05$, given by [0.037, 0.064].}
\label{fig: s3}
\end{figure}

Figure~\ref{fig: s3} shows that the type I error of the random shift methods increases as $\theta \to 0$. This is more evident for the dependent and confounded covariate $x_4(\mathbf{s})$. In contrast, the power for $x_1(\mathbf{s})$ and $x_2(\mathbf{s})$ remains stable and comparable to that of the classical method. These findings support the default choice $\theta = 1$, which achieves nominal level control while maintaining competitive power.

\clearpage
\section{Additional Figures for the Real Data Application}

\begin{figure}[htbp]
\centerline{\includegraphics[width=\textwidth]{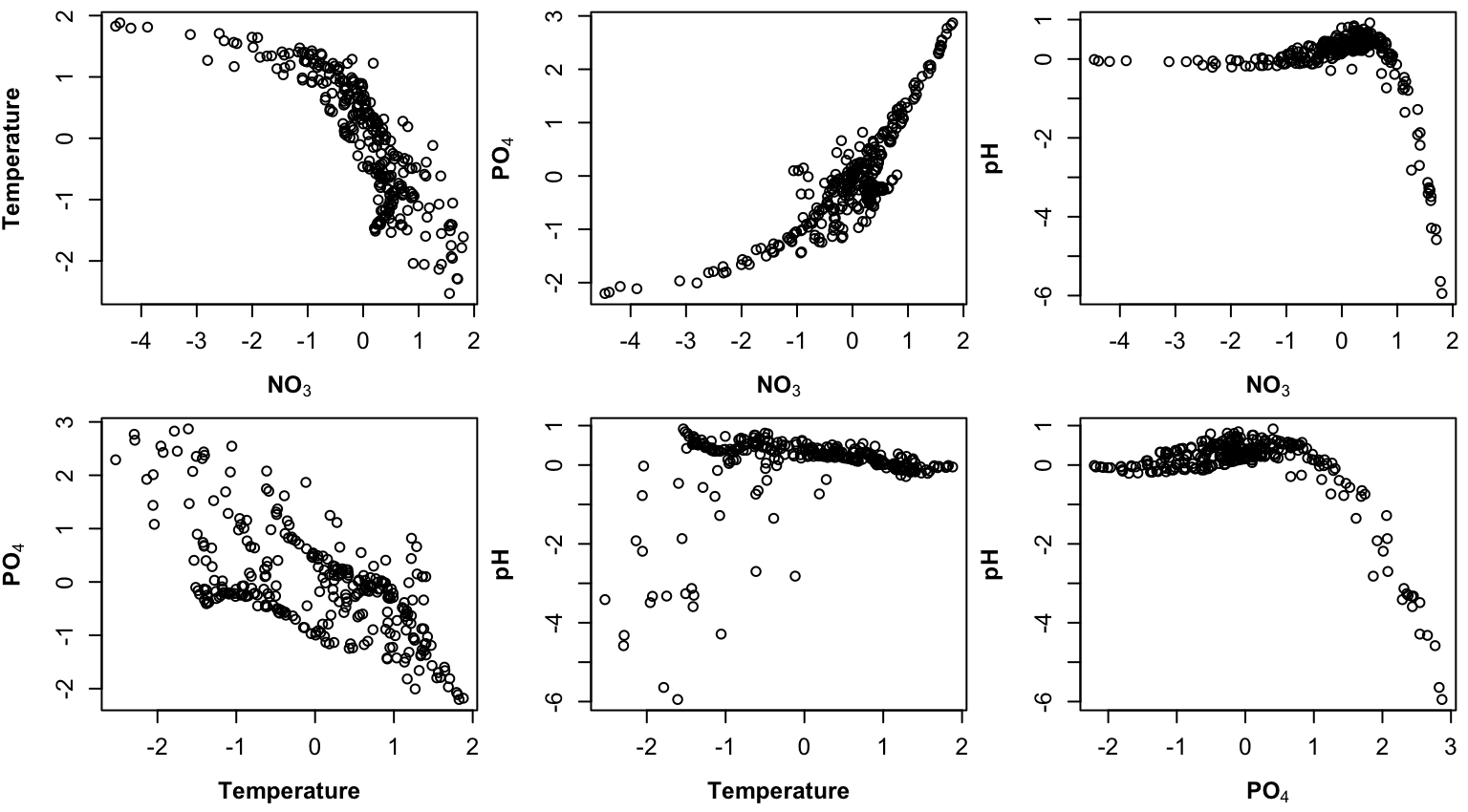}}
\caption{Scatter plots of the covariates.}
\label{plot_betweencov}
\end{figure}

\bibliography{ref}

\end{document}